\documentclass[a4paper,12pt]{article} 

\usepackage[english]{babel}
\usepackage[T1]{fontenc} 
 \usepackage[utf8]{inputenc}
\usepackage{indentfirst}
\usepackage{yfonts}
\usepackage{fancyhdr}
\usepackage[pdftex]{graphicx}
\usepackage{amssymb}
\usepackage{amsmath} 
\usepackage{latexsym}
\usepackage{amsthm} 
\usepackage{eucal} 
\usepackage{amsthm}
\usepackage{amstext}
\usepackage{dcpic, pictex}
\usepackage{bbold}
\usepackage{soul,color}

\theoremstyle{plain} 
\newtheorem{thm}{Theorem}[section] 

\newtheorem{corollary}[thm]{Corollary} 
\newtheorem{lem}[thm]{Lemma}

\theoremstyle{definition} 
\newtheorem{defn}{Definition}

\theoremstyle{remark} 
\newtheorem{oss}{Remark} 
\theoremstyle{remark} 
\newtheorem{remark}{Remark}

\usepackage{bookmark}
\usepackage{hyperref}
\def\tk{{{\tt k}}}
\def\cI{{\cal I}}
\def\cJ{{\cal J}}
\def\cR{{\cal R}}

\title{A large probability averaging Theorem for the defocousing NLS}
\author{D. Bambusi\footnote{Dipartimento di Matematica, Universit\`a degli Studi di Milano, Via Saldini 50, I-20133
Milano. \newline
\textit{Email: } \texttt{dario.bambusi@unimi.it}},
A. Maiocchi\footnote{Dipartimento di Matematica, Universit\`a degli Studi di Milano, Via Saldini 50, I-20133
Milano. \newline
\textit{Email: } \texttt{alberto.maiocchi@unimi.it}},
L. Turri\footnote{Dipartimento di Matematica, Universit\`a degli Studi di Milano, Via Saldini 50, I-20133
Milano. \newline
\textit{Email: } \texttt{luca.turri@unimi.it}}}

\begin{document}
\maketitle
\begin{abstract}
We consider the nonlinear Schr\"odinger equation on the
one dimensional torus, with a defocousing polynomial nonlinearity and
study the dynamics corresponding to initial data in a set of large
measure with respect to the Gibbs measure. We prove that along the
corresponding solutions the modulus of the Fourier coefficients is
approximately constant for times of order $\beta^{2+\varsigma}$,
$\beta$ being the inverse of the temperature and $\varsigma$ a
positive number (we prove $\varsigma= 1/10$). The proof is obtained by adapting
to the context of Gibbs measure for PDEs some tools of Hamiltonian
perturbation theory.
\end{abstract}
\section{Introduction and statement of the main result. }
In this paper we study the dynamics of the defocusing NLS with a
polynomial nonlinearity. We show that, with large
probability in the sense of Gibbs measure, each of the
actions of the unperturbed system is approximately invariant for long
times. This is obtained by generalizing to the context of PDEs some
tools of perturbation theory in Gibbs measure developed in recent
years in the context of lattice dynamics
\cite{Car07,CM12,BCM14,DeRoHuv15,BCMM15}.

The system we consider is the defocousing NLS on the torus
\begin{equation}\label{schreq}
i\dot{\psi}=-\Delta\psi+F'\left(|\psi|^2\right)\psi,\;\;
x\in\mathbb{T},
\end{equation}
where $F$ is a polynomial of degree $q\geq 2$,
$F(x):=\sum_{j=2}^qc_jx^j$, s.t. $F(x)\geq 0$ for any $x\geq0$ and
$c_2\not=0$. The flow of \eqref{schreq} is almost surely globally well-posed on any one of the spaces $H^s$ with $s$ fulfilling
$\frac{1}{2}-\frac{1}{q-1}<s<\frac{1}{2}$ (see
e.g. \cite{Bourgain94,BGT01}, see also \cite{ColOh}). \textit{We fix $s$ in this range once for all}.

We recall that  the Gibbs measure is  formally defined by
\begin{align}
\label{gibbs}
\mu_{\beta}=\frac{e^{-\beta
    \left(H(\psi)+\frac{1}{2}\|\psi\|^2_{L^2}\right)}}{Z(\beta)},\;
\beta>0\ ,\quad Z(\beta):=\int_{H^s}e^{-\beta
  \left(H(\psi)+\frac{1}{2}\|\psi\|^2_{L^2}\right) }d\psi
d\bar{\psi}\ ,
\end{align}
where $H$ is the Hamiltonian of the NLS (see \eqref{ham}) and $\beta$
plays the role of the inverse of the temperature {(we add the
  $L^2$-norm to avoid the frequency $0$).} We study the system in the
limit of $\beta$ large.\\ We denote by $\psi_\tk$ the ${\tt k}-$th
Fourier coefficients of $\psi$ defined by
$\psi_{\tk}:=\frac{1}{\sqrt{2\pi}}\int_0^{2\pi}\psi(x)e^{-i\tk x}dx $.

Our main result is the
following one
\begin{thm}\label{consaz}
There exist  $\beta^*, C,C'>0$ s.t. for any
$\eta_1,\eta_2>0$, any $\beta$ fulfilling
$$
\beta>\max\left\{\beta_*,\frac{C}{\eta_1^{\frac{10}{7}}\eta_2^{\frac{5}{7}}} \right\}
$$
and any $\tk\in\mathbb{Z}$, there
exists a measurable set $\cJ_\tk\subset H^s$ whose complement
$\cJ_\tk^c$ has small measure, namely 
$\mu_{\beta}(\cJ_\tk^c)<\eta_2$ 
s.t., if the initial datum
$\psi(0)\in\cJ_{\tk}$ then the solution exists globally in $H^s$ and
one has 
\begin{equation}\label{cambioazione}
  \left|
\frac{|\psi_{\tt k}(t)|^2-|\psi_{\tt
  k}(0)|^2}{{C'}/{{(1+{\tt k}^2)\beta}}}
  \right|<\eta_1
\ ,\;\; \forall
|t|<C'\eta_1\sqrt{\eta_2}\beta^{2+\varsigma}\ ,\ \varsigma=\frac{1}{10}\ .
\end{equation}
\end{thm} 
\begin{oss}
The expectation value of  $\psi_\tk$ is
$C_1/\sqrt{(1+{\tt k}^2)\beta}$,
with a suitable constant $C_1$.
\end{oss}
\begin{corollary}
  \label{tutte}
Under the same assumption of Theorem \ref{consaz} and for any $\alpha<1/2$, there
exists a measurable set $\cI_\alpha\subset H^s$ with
$\mu_{\beta}(\cI_\alpha^c)<\eta_2$ 
s.t., if the initial datum
$\psi(0)\in\cI_{\alpha}$ then the solution exists globally in $H^s$
and one has 
\begin{equation}
\label{tutte.1}
\left|\frac{|\psi_{\tt k}(t)|^2-|\psi_{\tt
  k}(0)|^2}{\left[(1+\tk^2)^{\alpha}\beta\right]^{-1}}\right|<\eta_1
  \;\;\ ,\;\; \forall
|t|<C'\eta_1\sqrt{\eta_2}\beta^{2+\varsigma}\ \ ,\quad \forall \tk\in\mathbb{Z}.
\end{equation}
\end{corollary}

\begin{remark}
\label{actions}
The quantity $|\psi_{{\tt k}}|^2$ appears since it is the action
of the linearized system. Theorem \ref{consaz} shows that, for general
initial data, $|\psi_{{\tt k}}|^2$ moves very little compared to its typical size over a
time scale of order $\beta^{2+\varsigma}$. Corollary \ref{tutte}
controls all the actions at the same time at the prize of giving a slightly worst
control on the actions with large index.
\end{remark}

\begin{remark}
\label{time}
{If one considers \eqref{schreq} as a perturbation of the cubic
integrable NLS, then one has that the main term of the perturbation is
(in the equation) $\left|\psi\right|^4\psi$ whose size can be thought
to be of order $\beta^{-5/2}$ which is of order $\beta^{-2}$ smaller
then the linear part. For this
reason one can think that the effective perturbation is of size 
$\beta^{-2}$. So one expects to obtain a control of the dynamics of
the actions over a time scale of order $\beta^2$.}

Theorem \ref{consaz}, not only gives a rigorous proof of this fact,
but also shows that the actions remain approximatively constant over a
longer time scale.  We do not expect the value of $\varsigma$ to be
optimal.
 \end{remark}

\begin{remark}
\label{small den}
In order to cover times longer than $\beta^{-2}$, we have to face the
problem of small denominators. Indeed, over the longer time scale, the
nonlinear corrections to the frequencies become relevant and the heart
of the proof consists in giving an estimate of the measure of the
phase space in which the nonlinear frequencies are nonresonant.
\end{remark}

\vskip 10pt

Theorem \ref{consaz} is essentially an averaging theorem for
perturbations of a linear resonant system.  

We recall that previous results giving long time stability of the
actions in \eqref{schreq} have been obtained in \cite{Bam99} and \cite{Bourgain00}. The
first two results allow to control the dynamics for exponentially long
times, but only for initial data close in energy norm to some finite dimensional
manifold, so essentially for a very
particular set of initial data. Bourgain \cite{Bourgain00} was able to
exploit the nonlinear modulation of the frequencies in order to show
that for most (in a suitable sense, not related to Gibbs measure)
initial data in $H^s$ with $s\gg1$ the Sobolev norm of the solution is
controlled for times longer then any inverse power of the small
parameter.

Nothing is known for solutions with low regularity as those dealt with
in the present paper. 

Our result can be compared also to the result of Huang Guan \cite{Huang13},
who proved a large probability averaging theorem for perturbations of
KdV equation. We emphasize that the result of \cite{Huang13}
deals with the quite
artificial case in which the perturbation is smoothing, namely it maps
functions with some regularity into functions with higher
regularity. In our case we deal with the natural local perturbation
given by a polynomial in $\psi$. Furthermore \cite{Huang13} only deals with
smooth solution. We also recall  \cite{HKM} in which a weaker version
of averaging theorem is obtained for solutions of some NLS-type equations. In that paper the initial datum is required to be more regular that in Theorem \ref{consaz} and the times covered are shorter.

Finally we mention the papers \cite{BDGS,BG,Bam03} which deal with
very smooth initial data and perturbations of nonresonant linear
system. These results are clearly in a context very different from
ours.
\vskip 10pt
The proof of our result is based on the generalization to the context
of Gibbs measure for PDEs of Poincar\'e's method of construction of
approximate integrals of motion \cite{poincare,GIOGGAL78}. The standard
way of using this method consists in first using a formal algorithm
giving the construction of objects which are expected to be
approximate integrals of motion and then adding estimates in order to
show that this  actually happens. This is  the way we proceed. 

So, first, we develop a formal scheme of construction of the
approximate integrals of motion which is slightly different from the
standard one. This is due to the fact that the linearized system is
completely resonant and we have to find a way to use the nonlinear
modulation of the frequencies in order to control each one of the
actions. We have also to restrict our construction to the region of
the phase space in which the frequencies are nonresonant. This is
obtained by eliminating (through cutoff functions) the regions of the
phase space where the linear combinations of the frequencies that are
met along the construction are smaller than $\delta$, where $\delta$
is a parameter that will be determined at the end of the construction.

The formal construction is contained in Sect.~\ref{formalconstr}. As a
result of this section, for any $\tt k$, we obtain a function
$\Phi_{\tt k}(\psi)$ close to $\left|\psi_{\tk}\right|^2$ which is
expected to be an approximate integral of motion.


The second step of the proof consists in estimating the $L^2(\mu_{\beta})$ norm of
$\dot \Phi_{\tt k}$ and in showing that it is small. To this end, we first recall that all the
estimates can be done by working with the Gaussian measure associated
to the linearized system, then we introduce the class of functions
which will be needed for the construction.  Then we show how  to control the $L^2(\mu_{\beta})$ norm
of such functions. This is obtained by exploiting
the decay of the Fourier modes of functions in the support of the Gibbs measure. Then we
use similar ideas in order to show that the integral of a function of
our class on the resonant region is small with $\delta$. Then we
choose $\delta$ in order to minimize the $L^2(\mu_{\beta})$ norm of $\dot
\Phi_{\tt k}$. 

Finally we use the invariance of the Gibbs measure and Chebyshev
theorem in order to pass from the estimate of $\dot
\Phi_{\tt k}$ to the estimate of $|\Phi_{\tt k}(t)-\Phi_{\tt k}(0)|$.

\noindent
{\it Acknowledgements.} {We thank T. Oh and N. Burq for introducing us to the
theory of Gibbs measure for PDEs.}

\section{Preliminaries}
Explicitly,  the Hamiltonian of \eqref{schreq}  is given by 
\begin{equation}\label{ham}
H=H_2+P\
\end{equation}
where
$$H_2:=\frac{1}{2}\int_0^{2\pi}|\nabla\psi(x)|^{2}dx,$$
$$ P=\sum_{j=2}^qH_{2j}, \;\; H_{2j}:=\frac{c_j}{2j}\int_0^{2\pi}|\psi(x)|^{2j}dx.$$
We will denote by $\Phi_{NLS}^t$ its flow (see \cite{BGT04}).
We consider the Gibbs measure $\mu_{\beta}$ associated to this Hamiltonian, which
is known to be invariant with respect to $\Phi_{NLS}^t$
(\cite{Bourgain94,LRS88,Tzv08,TT10}).

Given a function $f:H^s\rightarrow \mathbb{C}$, $f\in L^2(H^s,\mu_{\beta})$, we define its average and its $L^2$-norm with respect to the measure $\mu_{\beta}$ as:
$$\left<f\right>:=\int_{H^s}fd\mu_{\beta}$$
$$\|f\|_{\mu_{\beta}}^2:=\int_{H^s}|f|^2d\mu_{\beta}$$
\begin{oss}
From the invariance of $\mu_{\beta}$, one has that  the average $\left<f\right>$ and the $L^2$-norm $\|f\|_{\mu_{\beta}}$ of  the functions are preserved along the flow, namely $\left<f\circ \Phi^t_{NLS} \right>=\left<f\right>,$ $\|f\circ \Phi^t_{NLS} \|_{\mu_{\beta}}=\|f\|_{\mu_{\beta}}$ for any $t$.
\end{oss}

From now on, we shall work using the Fourier coordinates.
In these coordinates, $H_2$ becomes 
$$ H_2:=\frac{1}{2}\sum_k k^2|\psi_k|^2.$$

We give now some results on the relationship of the Gaussian measure
with the Gibbs measure. Define the $H^1$-norm:
$$\|\psi\|_{H^1}^2:=\sum_k(1+k^2)|\psi_k|^2,$$
then we can express  $H_2+\|\psi\|_{L^2}^2=\frac{1}{2}\|\psi\|_{H^1}^2$ and
the Gaussian measure is formally defined by
\begin{equation}
\mu_{g,\beta}:=\frac{e^{-\frac{\beta}{2}\|\psi\|_{H^1}^2}}{Z_g(\beta)},
\end{equation}
with $$ Z_g(\beta):=\int_{H^s}e^{-\frac{\beta}{2}\|\psi\|_{H^1}^2}d\psi d\bar{\psi}.$$
Given a function $f:H^s\rightarrow \mathbb{C}$, we denote by $$\|f\|_{g,\beta}^2:=\int_{H^s}|f|^2d\mu_{g,\beta}$$ its $L^2$-norm respect to $\mu_{g,\beta}$.

The following lemmas will be proved in Appendix \ref{sezionemisura}.
\begin{lem}\label{gibbsgauss}
 There exist $\beta^*,\tilde{C}>0$ s.t. for any $\beta>\beta^*$ and for any function $f\in L^2(H^s,\mu_{g,\beta})$, one  has:
$$\|f\|_{\mu_{\beta}}\leq\|f\|_{g,\beta} e^{\tilde{C}}.$$
\end{lem}
We emphasize that the constant $\tilde{C}$ is independent of $\beta$ and $q$.

\begin{lem}\label{stimabasso}
There exists $C_{sob},D'>0$ s.t. for any $\beta>0$ and any function $f\in L^2(H^s,\mu_{g,\beta})$, one has
$$\left\|f\right \|_{\mu_{\beta}}\geq e^{-\frac{C_{sob}}{2\beta}q\max_j{c_j}D'^j}\left\|f\chi_{\left\{\|\psi\|_{H^{s_1}}<\frac{D'}{\beta}\right\}}\right\|_{g,\beta}$$
where $\chi_{\{U\}}(\psi)$ is the characteristic function of the set $U$.
\end{lem}

The next lemma shows that every moment of $\mu_{\beta}$ is well defined.
\begin{lem}\label{momenti gaussiana}
There exists $\beta^*>0$ s.t., for any $s_1<\frac12$, $n\in\mathbb{N}$,  $\beta>\beta^*$, one has $$\|\psi\|_{H^{s_1}}^n\in L^1(H^{s},\mu_{\beta})\cap L^1(H^{s},\mu_{g,\beta}).$$
\end{lem}
Finally, for the special case of the function $|\psi_{\tt k}|^2$, we have the following 
lemma. 
\begin{lem}\label{stimaazione}
 There exists $\beta^*>0,C>0$ s.t. for any $\beta>\beta^*$ s.t. $$\left\||\psi_{\tt k}|^2\right \|_{\mu_{\beta}}\geq\frac{C}{\beta \left(1+{\tt k}^2\right)}.$$ 
\end{lem}

\section{Polynomials with frequency dependent coefficients}\label{classifunzioni}
In this section we introduce a class of function on $H^s$ which will be stable under the perturbative construction and we prove some results needed for the rest of the proof.
\begin{defn}\label{polinomio}
Let $B_1,B_2$ be  two Banach spaces, we say that $F(y):B_1\rightarrow B_2$ is a polynomial of degree $n$ if there exists a $n$-multilinear form $\tilde{F}$ s.t. for any $y\in B_1$, one has $F(y)= \tilde{F}(\underbrace{y,y,...,y}_n)$.
\end{defn}
\begin{oss}
In particular a polynomial $f:H^s\rightarrow \mathbb{C}$ of degree $n$ has the form:
\begin{equation}\label{poli_n}
f(\psi)=\sum_{l,m}\psi^l\bar{\psi}^mf_{l,m}
\end{equation}
where $l=\{l_k\}$,  $m=\{m_k\}$, $l_k,m_k\in\mathbb{N}$, $\sum_k l_k+m_k=n$, $f_{l,m}\in \mathbb{C}$, $\psi^l=...\psi_{-k}^{l_{-k}}...\psi_{k}^{l_{k}}...$ and the  same for $\bar{\psi}^m$.
\end{oss}
\begin{defn}
We say that a polynomial $f$ of the form $\eqref{poli_n}$ of degree $2n$ is of class $P_{2n}$ if it fulfills the \textit{null momentum} condition, i.e. 
\begin{equation}\label{momento_nullo}
f_{l,m}\neq 0\mbox{ only if }\sum_{k\in Supp(l)} k=\sum_{k\in Supp(m)} k \mbox{ and } \sum_kl_k=\sum_km_k=n.
\end{equation} 
On $P_{2n}$, we  introduce the following norm 
\begin{equation}\label{3sbarre}|||f|||:=\sup_{l,m}\left|f_{l,m}\right|.
\end{equation}
\end{defn}
\begin{oss}
In the following, due to \eqref{momento_nullo},  we will write a polynomial $f\in P_{2n}$  also in the equivalent following form, more convenient in a lot of situations
\begin{equation}\label{poli_n_nuovo}
f(\psi)=\sum_{\substack{k=(k_1,...,k_{2n})\\ \sum_{i=1}^nk_i=\sum_{i=n+1}^{2n}k_i}}f_k\prod_{i=1}^{n}\psi_{k_i}\bar{\psi}_{k_{i+n}}.
\end{equation}
\end{oss}
The next lemma shows that the polynomials of class $P_{2n}$ are smooth polynomials on $H^{s_1}$, $\frac12-\frac1n<s_1<\frac12$. 
\begin{lem}\label{convolution}
Let $n$ be a positive integer and $s_1$ s.t. $\frac{1}{2}-\frac{1}{2n}<s_1<\frac12$, $f\in P_{2n}$, then there exists $C(s_1,n)>0$ s.t. 
\begin{equation}|f(\psi)|\leq C(s_1,n)\|\psi\|_{H^{s_1}}^{2n}|||f|||.
\end{equation}
\end{lem} 
\proof
$$|f(\psi)|\leq\sum_{\substack{k_1,...,k_{2n}\\ \sum_{i=1}^nk_i=\sum_{i=n+1}^{2n}k_i}}|f_{k_1,...,k_{2n}}|\prod_{i=1}^{2n}|\psi_{k_i}|$$
$$\leq |||f|||\sum_{\substack{k_1,...,k_{2n}\\ \sum_{i=1}^nk_i=\sum_{i=n+1}^{2n}k_i}}\prod_{i=1}^{2n}|\psi_{k_i}|$$
We define $\varphi:=\left\{\varphi_k\right\}:=\left\{|\psi_k|\right\}$, so, using Sobolev's embedding $H^{s_1}\subset L^{2n}$ for $\frac{1}{2}-\frac{1}{2n}<s_1<\frac12$, one has: 
$$|f(\psi)|\leq |||f|||\sum_{\substack{k_1,...,k_{2n}\\ \sum_{i=1}^{n}k_i=\sum_{i=n+1}^{2n}k_i}}\prod_{i=1}^{2n}\varphi_{k_i}=\|\varphi\|^{2n}_{L^{2n}}|||f|||$$
$$\leq{C}(s_1,n)\|\varphi\|^{2n}_{H^{s_1}}|||f|||={C}(s_1,n)\|\psi\|^{2n}_{H^{s_1}}|||f|||.$$
\endproof

We will also consider the functions $f\in C^r(\ell^1,P_{2n})$, $f:\ell^1\ni\omega=\left\{\omega_j\right\}\rightarrow f(\psi,\omega)=\sum_{\substack{k=(k_1,...,k_{2n})\\ \sum_{i=1}^nk_i=\sum_{i=n+1}^{2n}k_i}}f_k(\omega)\prod_{i=1}^{n}\psi_{k_i}\bar{\psi}_{k_{i+n}}$. In the following $\omega_j$ will be the nonlinear modulation of the $j$-th frequency.\\
Actually we need to keep the information of the size of the different derivative of $f$.  
So, we give the following definition.
\begin{defn}\label{parte polinomio e parte no}
We will say that 
$f\in P^r(2n,\{A_i\}_{i=0}^r)$ 
 if
 $f\in C^r(\ell^1,P_{2n})$ and $$\sup_{\substack{\omega, k\\|j|=i}}\left|\frac{\partial^{|j|}f_k(\omega)}{\partial {\mathbf\omega}^j}\right|<A_i, \;\; \forall i=0,...,r .$$
\end{defn} 
\begin{oss}
$\mbox{Max}_i{A_i}$ is a norm for $C^r(\ell^1,P_{2n})$.
\end{oss}
Given a function $f\in C^r(\ell^1,P_{2n})$,  we also consider $$f_{ph}(\psi):=f(\psi,|\psi|^2),$$ conversely, we will say that $\tilde{f}:H^s\rightarrow \mathbb{C}$ is of class $P^r(2n,\{A_i\}_{i=0}^r)$  if there exists a function $F(\psi,\omega)\in P^r(2n,\{A_i\}_{i=0}^r)$ s.t. 
$F(\psi,\omega)_{|_{\omega=\{|\psi_k|^2\}}}=\tilde{f}(\psi)$.

\begin{oss}
If $f\in P_{2n}$ with $|||f|||<\infty$,  then $f\in  P^{\infty}(2n,\{A_i\}_{i=0}^{\infty})$ with $A_0=|||f|||$ and $A_i=0$ for any $i\geq0$. For simplicity, we will write $f\in  P^{\infty}(2n,|||f|||)$.
\end{oss}

\begin{oss}
From Lemma $\ref{convolution}$,   for any $n\in \mathbb{N}$ and for any $s_1$ s.t. $\frac{1}{2}-\frac{1}{2n}<s_1<\frac12,$ for any $r\geq 0$ and for any $f\in P^r(2n,\{A_i\}_{i=0}^r)$,  one has
\begin{equation}
|f(\psi)|\leq A_0C(s_1,n)\|\psi\|_{H^{s_1}}^{2n}.
\end{equation}
\end{oss}
The connection of the norm of $ P^0(2n,A_0)$ and the $L^2$-norm is given by
\begin{lem}
\label{gausemplice}
Let $n$ be an integer, denoted by $C_g(n):=2^{n+2}[(2n)!]^{\frac{3}{2}}(2n-1)^2\left(\sum_l \frac{1}{1+l^2}\right)^{n},$ for any  $\beta>0$, and $f_{ph}\in P^0(2n,A_0),$ one has
\begin{equation}
\|f_{ph}\|_{g,\beta}\leq \frac{A_0C_g(n)}{\beta^{n}}.
\end{equation}
\end{lem}
\proof
Writing $f_{ph}=\sum_{\substack{k=(k_1,...,k_{2n}) }}f_{k}(\psi)\prod_{i=1}^n\psi_{k_i}\bar{\psi}_{k_{n+i}}$, one has

\begin{equation}\label{riducodominiointegrazione}\|f_{ph}\|^2_g=\int_{H^s}|f|^2d\mu_{g,\beta}=\int_{H^s}\sum_{\substack{k,j }}f_{k}(\psi)\bar{f}_{j}(\psi)\prod_{i=1}^n\psi_{k_i}\psi_{j_{n+i}}\bar{\psi}_{j_{i}}\bar{\psi}_{k_{n+i}}d\mu_{g,\beta}.
\end{equation}

Let $s_1$ be s.t. $\max\left\{s,\frac{n-1}{2n}\right\}<s_1<\frac{1}{2}$,  by Lemma $\ref{convolution}$, there exists a constant $C$ s.t. $|f|^2\leq CA_0^2\|\psi\|^{4n}_{H^{s_1}}$, moreover by Lemma $\ref{momenti gaussiana}$, $\|\psi\|^{4n}_{H^{s_1}}\in L^1(H^{s},\mu_{g,\beta})$.
So we can exchange the order between the integral and the series and $\eqref{riducodominiointegrazione}$ becomes
$$\sum_{\substack{k,j }}\int_{H^{s}}f_{k}(\psi)\bar{f}_{j}(\psi)\prod_{i=1}^n\psi_{k_i}\psi_{j_{n+i}}\bar{\psi}_{j_{i}}\bar{\psi}_{k_{n+i}}d\mu_{g,\beta}=$$
\begin{equation}\label{primadiintegrare}\sum_{k,j}\frac{\int_{H^{s}}f_{k}(\psi)\bar{f}_{j}(\psi)\prod_{i=1}^n\psi_{k_i}\psi_{j_{n+i}}\bar{\psi}_{j_{i}}\bar{\psi}_{k_{n+i}}e^{-\frac{\beta}{2}\sum_{ S_{kj}}\left(1+l^2\right)|\psi_l|^2}\prod_{ S_{kj}}d\psi_ld\bar{\psi}_l}{\prod_{ S_{kj}}\int_{H^s} e^{-\frac{\beta}{2}\left(1+l^2\right)|\psi_l|^2}d\psi_ld\bar{\psi}_l}
\end{equation}
where $S_{kj}:=Supp(k,j)$.  It is usefull to use the following
notation: given a set $K$ of indices $(k_1,...,k_{2n})$ with an even
number of components, we denote
$$
K_1:=\{k_1,...,k_n  \}\ ,\quad K_2:=\{k_{n+1},...,k_{2n}  \} \ . 
$$

Using the substitution
$\psi_l=\frac{\sqrt{2z_l}}{\sqrt{\beta(1+l^2)}}e^{i\theta_l}$, $z_l\in
\mathbb{R}^+,$ $\theta_l\in [0,2\pi)$, one has that the only integrals
  different from 0 are the terms in which $K_1\cup J_2=K_2\cup J_1$.

We denote by $\mathcal{T}$ the set of $(k,j)$ s.t. $K_1\cup
J_2=K_2\cup J_1$ and with both $k$ and $j$ fulfilling the zero
momentum condition, namely $\sum_{i=1}^n k_i=\sum_{i=n+1}^{2n}k_i $,
$\sum_{i=1}^n j_i=\sum_{i=n+1}^{2n}j_i$. Thus \eqref{primadiintegrare}
is bounded by
$$ A_0^2\sum_{\substack{k,j\in \mathcal{T} }}\frac{2^{2n}}{\beta^{2n}\prod_{i=1}^n\left(1+k^2_i\right)\left(1+j^2_{n+i}\right)}\int \prod_{i=1}^{n}z_{k_i}z_{j_{i+n}}e^{-\sum_{ S_{kj}}z_l}\prod_{ S_{kj}}dz_l$$
$$\leq A_0^2\frac{2^{2n}(2n)!}{\beta^{2n}}\sum_{\substack{k,j\in \mathcal{T} }}\frac{1}{\prod_{i=1}^n\left(1+k^2_i\right)\left(1+j^2_{n+i}\right)}.$$
So, 
\begin{equation}\label{serie}\|f_{ph}\|^2_g\leq\frac{A_0^22^{2n}(2n)!}{\beta^{2n}}\sum_{\substack{(k,j)\in \mathcal{T} }}\frac{1}{\prod_{i=1}^n\left(1+k^2_i\right)\left(1+j^2_{n+i}\right)}.
\end{equation}
Since we sum on $(k,j)\in \mathcal{T}$, we have that, having fixed
$K_1\cup J_2=K_2\cup J_1$ we have $(2n)!$ way to rearrange $K_1\cup
J_2$ and $(2n)!$ way to rearrange $K_2\cup J_1$,
so 
$$\sum_{\substack{(k,j)\in \mathcal{T}
}}\frac{1}{\prod_{i=1}^n\left(1+k^2_i\right)\left(1+j^2_{n+i}\right)}\leq
[(2n)!]^2\sum_{\substack{k_1,...,k_n,\\j_{n+1},...,j_{2n}}}\frac{1}{\prod_{i=1}^n\left(1+k^2_i\right)\left(1+j^2_{n+i}\right)}$$
$$= [(2n)!]^2\left(\sum_l \frac{1}{1+l^2}\right)^{2n}.$$
So, finally,
$$\|f_{ph}\|^2_g\leq\frac{A_0^22^{2n}[(2n)!]^{3}\left(\sum_i \frac{1}{1+i^2}\right)^{2n}}{\beta^{2n}}\leq\frac{A_0^2C^2_g(n)}{\beta^{2n}}$$ 
with $C_g(n)^2:=2^{2n+4}[(2n)!]^{3}(2n-1)^4\left(\sum_l \frac{1}{1+l^2}\right)^{2n}$.\\
\qed

\begin{oss}
According to Lemma $\ref{gibbsgauss}$, one also has 
\begin{equation}
\|f_{ph}\|_{\mu_{\beta}}\leq \frac{A_0C_g(n)}{\beta^{n}}.
\end{equation}
\end{oss}

The Poisson brackets of two functions $f,g$ with $f\in P_{2n}$ and $g\in  P^r \left(2m,\{A_i\}_{i=0}^r\right)$ its formally, given by
$$\left\{f,g\right\}:=L_f(g):=-i\sum_k\left(\frac{\partial f}{\partial \psi_k}\frac{\partial g}{\partial \bar{\psi}_k}-\frac{\partial g}{\partial \psi_k}\frac{\partial f}{\partial \bar{\psi}_k}\right).$$

\begin{oss}\label{n+m}
If $f\in P_n$, $g\in P_m$, then $$\left\{f,g\right\}\in P_{n+m-2}.$$
\end{oss}
\begin{lem}\label{parentesi_poisson_lemma}
Consider $f\in P_{2n}$, $|||f|||<D$, $g_{ph}\in P^r \left(2m,\{A_i\}_{i=0}^r\right)$.
Then
\begin{equation}\label{standard+fratto}\left\{f,g\right\}=F_1+F_2,\end{equation}
where
\begin{equation}\label{standard} F_1\in P^r \left(2n+2m-2,2nmD\{A_i\}_{i=0}^r\right),\end{equation}
\begin{equation}\label{fratto} F_2\in P^{r-1}(2n+2m,2nD\{A_{i+1}\}_{i=0}^{r-1}).\end{equation}
\end{lem}
\proof
Writing $g_{ph}=\sum_{k=(k_1,...k_{2m})}g_k\left(\left\{|\psi_k|^2\right\}\right)\psi_{k_1}...\psi_{k_m}\bar{\psi}_{k_{m+1}}...\bar{\psi}_{k_{2m}}$, then it is immediate to verify that $\eqref{standard+fratto}$ holds with

$$F_1=\sum_{k=(k_1,...k_{2m})}g_{k}(\left\{|\psi_j|^2\right\})\left\{f,\psi_{k_1}...\psi_{k_m}\bar{\psi}_{k_{m+1}}...\bar{\psi}_{k_{2m}}\right\}$$
$$F_2=\sum_{k=(k_1,...k_{2m})}\psi_{k_1}...\psi_{k_m}\bar{\psi}_{k_{m+1}}...\bar{\psi}_{k_{2m}}\left\{f,g_{k}(\left\{|\psi_j|^2\right\})\right\}=$$
$$=\sum_{k=(k_1,...k_{2m})}\left(\sum_l \frac{\partial g_{k}(\left\{|\psi_j|^2\right\})}{\partial \omega_l}\right)\psi_{k_1}...\psi_{k_m}\bar{\psi}_{k_{m+1}}...\bar{\psi}_{k_{2m}}\left\{f,|\psi_l|^2\right\}$$
and, by Remark \ref{n+m},  $F_1\in P^r \left(2n+2m-2,2nmD\{A_i\}_{i=0}^r\right)$ and $F_2\in P^{r-1}(2n+2m,2nD\{A_{i+1}\}_{i=0}^{r-1})$ hold.\\
\qed

Actually, we shall use a more particular class of functions in which the range of the indices is subject to  a further restriction. This is related to the fact that in our construction we shall fix an index $\tk$ corresponding to the action we want to conserve.
To this end, we introduce the following definition:

\begin{defn}Given $M>0,\; {\tt k}\in \mathbb{Z}$, a linear
  combination $$G{(k_1,...,k_{2n})}:=\sum_{i=1}^{2n}a_ik_i$$ with
  $a_i\in \mathbb{Z}$, $|a_i|\leq M$, we will say that the
  relation
$$G(k_1,...,k_{2n})=\tt k$$ is $(M,{\tt k})$-admissible.
\end{defn}

\begin{lem}\label{stima_par_poisson_1}
Given $D>0$, let be $f\in P_{2n}$, $|||f|||<D$, $g(\psi,\bar{\psi})\in  P^r(2m,\{A_i\}_{i=0}^r)$, $M>0$, ${\tt k}\in \mathbb{Z}$.\\
Assume that $$g=\sum_{\substack{k=(k_1,...,k_{2m}) \mbox{ s.t. }\\ G_k(k_1,...,k_{2m})=\tt k }}g_k\left(\left\{|\psi_k|^2\right\}\right)\psi_{k_1}...\psi_{k_m}\bar{\psi}_{k_{m+1}...}\bar{\psi}_{k_{2m}},$$
where, for any $k$, $G_k=\tt k$ is  $(M,{\tt k})$-admissible.
Then
$$\left\{f,g\right\}=F_1+F_2$$
where
\begin{align}
\label{F_1}F_1=\sum_{\substack{k'=(k'_1,...,k'_{2n+2m-2})\\\tilde{G}_{k'}(k'_1,...,k'_{2n+2m-2})=\tt k}}F_{1,k'}\psi_{k'_1}...\psi_{k'_{n+m-1}}\bar{\psi}_{k'_{n+m}...}\bar{\psi}_{k'_{2m+2n-2}}
\\
\label{F_2}F_2=\sum_{\substack{k''=(k''_1,...,k''_{2n+2m})\\\hat{G}_{k''}(k''_1,...,k''_{2n+2m})=\tt k}}F_{2,k''}\psi_{k''_1}...\psi_{k''_{m+n}}\bar{\psi}_{k''_{m+n+1}}...\bar{\psi}_{k''_{2m+2n}}
\end{align}

where for any $k',\; k''$, the relations $\tilde{G}_{k'}=\tt k$, $\hat{G}_{k''}=\tt k$ are $(2M,{\tt k})$-admissible.
\end{lem}
\proof
Writing $f=\sum_{\substack{l=(l_1,...l_{2n})
}}f_l\psi_{l_1}...\psi_{l_n}\bar{\psi}_{l_{n+1}...}\bar{\psi}_{l_{2n}}$,
by Lemma  \ref{parentesi_poisson_lemma}, we have 
$F_1\in P^r \left(2n+2m-2,2nmD\{A_i\}_{i=0}^r\right),\; F_2\in
P^{r-1}(2n+2m,2nD\{A_i\}_{i=1}^r)$.

Moreover, each term of $F_1$  is originated by two terms that depend respectively on $l=(l_1,...l_{2n})$ and $k=(k_1,...k_{2m})$ s.t. $\sum_{i=1}^nl_i=\sum_{i={n+1}}^{2n}l_i$,  $\sum_{i=1}^mk_i=\sum_{i={m+1}}^{2m}k_i$ and $\left\{l_1,...l_{n}\right\}\cap \left\{k_{m+1},...k_{2m}\right\}\neq \varnothing$
or $\left\{l_{n+1},...l_{2n}\right\}\cap \left\{k_{1},...k_{m}\right\}\neq \varnothing$.
Without losing generality,  we can suppose  $l_1=k_{m+1}$.\\
We form a vector of indices $k'=(l_2,...l_n,k_1,...,k_m,l_{n+1},...l_{2n}, k_{m+2},...,k_{2m})$ s.t. $\sum_{i=2}^nl_i+\sum_{i=1}^mk_i=\sum_{i={n+1}}^{2n}k_i+\sum_{i={m+2}}^{2m}k_i$.
Moreover, $k_{m+1}=\sum_{i=1}^mk_i-\sum_{i={m+2}}^{2m}k_i$.
By hypothesis, we can write $G_k(k_1,...,k_{2m})=\sum_{i=1}^{2m}a_ik_i$ with $a_i\in \mathbb{N}, |a_i|<M$,   so $${\tt k}=G_k(k_1,...,k_{2m})=\sum_{i=1}^{2m}a_ik_i=\sum_{i=1}^{m}(a_i+a_{m+1})k_i+\sum_{i=m+2}^{2m}(a_i-a_{m+1})k_i=$$
$$=\sum_{i=1}^{m}b_ik_i+\sum_{i=m+2}^{2m}b_ik_i=\tilde{G}_k(k_1,...,k_{m},k_{m+2},...,k_{2m})$$
$$=\tilde{G}_{k'}(l_2,...,l_n,k_1,...,k_{m},l_{n+1},...,l_{2n},k_{m+2},...,k_{2m}).$$
We note that $|b_i|<2M$ and $\tilde{G}_k$ is a linear combination only of $\{k_1,...,k_{m},k_{m+2},...,k_{2m}\}$ so it is independent of the \textit{null-momentum} condition related to \\$(l_2,...,l_n,k_1,...,k_{m},l_{n+1},...,l_{2n},k_{m+2},...,k_{2m})$, so we obtain the thesis for $F_1$.
For $F_2$ the situation is simpler.
Again   each term of $F_2$  is originated by two terms that depend respectively on $l$ and $k$ s.t. $\sum_{i=1}^nl_i=\sum_{i={n+1}}^{2n}l_i$,  $\sum_{i=1}^mk_i=\sum_{i={m+1}}^{2m}k_i$ and $\left\{l_1,...l_{n}\right\}\cap \left\{k_{m+1},...k_{2m}\right\}\neq \emptyset$
or $\left\{l_{n+1},...l_{2n}\right\}\cap \left\{k_{1},...k_{m}\right\}\neq \emptyset$.\\
We obtain a vector of indices $k''=(l_1,...,l_n,k_1,...,k_m,l_{n+1},...,l_{2n}, k_{m+1},...,k_{2m})$ s.t. $\sum_{i=1}^nl_i+\sum_{i=1}^mk_i=\sum_{i={n+1}}^{2n}k_i+\sum_{i={m+1}}^{2m}k_i$ and $${\tt k}=G_k(k_1,...k_{2m})=\tilde{G}_{k''}(l_1,...,l_n,k_1,...,k_{m},l_{n+1},...,l_{2n},k_{m+1},...,k_{2m}).$$
\qed
\begin{oss}This result holds also in the particular case in which $g_k$ is a constant independent of $\left\{|\psi_j|^2\right\}$.
 \end{oss}
In particular, one can obtain the following improvement of Lemma \ref{gausemplice}:
\begin{lem}\label{gau}
Let $n$ be an integer, $M>0$, ${\tt k}\in \mathbb{Z}$, let $$f_{ph}=\sum_{\substack{k=(k_1,...,k_{2n}) \\G_k(k_1,...,k_{2n})={\tt k}}}f_k\left(\{|\psi_k|^2\}\right)\psi_{k_1}...\psi_{k_n}\bar{\psi}_{k_{n+1}...}\bar{\psi}_{k_{2n}},$$ s.t. $f_{ph}\in P^0(2n,A_0)$ and  for any $k$,  $G_k(k_1,...,k_{2n})=\tt k$ is $(M,{\tt k})$-admissible. \\
Then, for any $\beta>0$, one has
 \begin{equation}\|f_{ph}\|_{g,\beta}\leq \frac{A_0C_g(n)M^2}{\left(1+{\tt k}^2\right)\beta^{n}}.\end{equation}
\end{lem}
The proof of this lemma is very technical and it is  deferred to Appendix \ref{Gaussian estimates}.

\section{Formal construction of perturbed actions}\label{formalconstr} 

In this section we look for a formal integral of motion which is a
higher order perturbation of $\Phi_{\tk,2}:= |\psi_{\tk}|^2$. Thus we
fix once for all the value of $\tk$.

To present the construction, we describe first an equivalent one, which however is difficult to manage directly.
Since $H_2$ is completely resonant, it is well known that one can
construct, formally a canonical transformation $T$ which transforms the
Hamiltonian into
\begin{equation}
\label{z6}
H_2+Z_4+Z_6+R_8
\end{equation}
with $Z_4$ and $Z_6$ which Poisson commute with $H_2$. In particular
$Z_4$ has been computed in many papers (see e.g. \cite{Bam99}  ) and
is given by 
\begin{equation}
\label{z4}
Z_4(\psi):=\frac{c_2}{2}\left(\sum_{k}|\psi_{k}|^2\right)^2-\frac{c_2}{2}\sum_{k}|\psi_{k}|^4\ .
\end{equation}

Then, following the ideas by Poincar\'e, we look for $\tilde
\Phi_{\tk,6}$, Poisson commuting with $H_2$,
s.t. $\tilde\Phi^{(6)}_{\tk}:=\Phi_{\tk,2}+\tilde\Phi_{\tk,6}$ is an approximate
integral of motion of \eqref{z6}. Computing the Poisson bracket of
this quantity with \eqref{z6}, one has that this is a quantity of
order at least 8 if
\begin{equation}
\label{n.8}
\left\{Z_4, \tilde \Phi_{\tk,6} \right\}=\left\{\Phi_{\tk,2},
Z_6\right\}=:\cR_6\ ,
\end{equation} 
which is clearly impossible since the l.h.s. is of order 8 and the r.h.s. of order 6, so we will modify it. Since $Z_4$ depends on the actions only,
one has 
$$ \left\{Z_4, \cdot \right\}=i\sum_{j}
\omega_j\left({\psi}_j\frac{\partial}{\partial \psi_j}-
\bar\psi_j\frac{\partial}{\partial \bar\psi_j}\right)\ ,
$$
with $\omega_j:=c_2\left(|\psi_j|^2+\sum_k|\psi_k|^2\right)$. So one is led to separate the regions
where the $\omega_j$'s are resonant and those in which they are non
resonant. The resonant regions and the nonresonant regions will be
defined precisely in the following. Denote $\cR_6^{NR}$ the restriction
of $Z_6$ to the nonresonant regions, we will solve the equation 
\begin{equation}
\label{nonres}
\left\{Z_4, \tilde \Phi_{\tk,6}  \right\}=\cR_6^{NR}\ .
\end{equation}
Looking for $\tilde
\Phi^{(6)}_{\tk}$ in the class of polynomials with frequency dependent coefficients, the approximate integral of motion that we are going to construct is
given by the sixth order truncation of $T^{-1}\tilde
\Phi^{(6)}_{\tk}$.
We proceed now to the construction of the  integral of motion.
Define the operator $L_{H_2}:=\left\{H_2,\cdot\right\}$, we have that for any $f\in P_{2n}$
$$L_{H_2}f=\left\{H_2,f\right\}\equiv-i\sum_{l,m}f_{l,m}\left<\mathbf{k^2}, (l-m) \right>\psi^l\bar{\psi}^m$$ 
where $\left<\mathbf{k^2}, (l-m) \right>:=\sum_j k_j^2(l_j-m_j)$.\\
Equivalently, for any for any $f\in P_{2n}$, we can write  
$$L_{H_2}f=-i\sum_{k}f_{k}\left(\sum_kk^2\left(\sum_{i=1}^n\delta_{k_i,k}-\sum_{i=n+1}^{2n}\delta_{k_i,k}\right)\right)\prod_{i=1}^n\psi_{k_i}\bar{\psi}_{k_{i+n}},$$
where $\delta_{x,y}$ is  kronecker's delta.
\begin{defn}\label{ker,range}
We denote by $$N_{H_2}:=\mbox{ker}L_{H_2}=\left\{f\in \cup_{n\in \mathbb{N}}P_{2n}: f_{l,m}\neq 0\Leftrightarrow\left<\mathbf{k^2}, (l-m) \right>=0\right\},$$ 
$$R_{H_2}:=\left\{f\in \cup_{n\in \mathbb{N}}P_{2n}: f_{l,m}\neq 0\Leftrightarrow\left<\mathbf{k^2}, (l-m) \right>\neq0\right\}.$$
\begin{oss}
$L_{H_2}:R_{H_2}\rightarrow R_{H_2}$ is formally invertible. 
\end{oss}
Given a polynomial $f$, we
indicate the projection of $f$ on $N_{H_2}$ by $f^{N_{H_2}}$ and the
projection on $R_{H_2}$ by $f^{R_{H_2}}$.
\end{defn}
In particular, we have
$$H_4^{R_{H_2}}:=\frac{c_2}{4}\sum_{\substack{k_1+k_2=k_3+k_4\\k^2_1+k^2_2\neq
    k^2_3+k^2_4}}\psi_{k_1}\psi_{k_2}\bar{\psi}_{k_3}\bar{\psi}_{k_4},$$ 
$$Z_4=H_4^{N_{H_2}}$$
Define now 
 $$\chi_4:=-L_{H_2}^{-1}H_4^{R_{H_2}},\;\chi_6:=-L_{H_2}^{-1}\left(\frac{1}{2}\left\{\chi_4,H_4^{R_{H_2}}\right\}+ \left\{\chi_4,Z_4\right\}+H_6\right)^{R_{H_2}},$$
$$\Phi_{k,4}:=L_{\chi_4}|\psi_k|^2,\;\Phi_{k,6}:=\frac{1}{2}L^2_{\chi_4}|\psi_k|^2+L_{\chi_6}|\psi_k|^2$$
and
$$
Z_6:=H_6^{N_{H_2}}+\left(\frac{1}{2}\left\{\chi_4,H_4^{R_{H_2}}\right\}+ \left\{\chi_4,Z_4\right\}\right)^{N_{H_2}},
$$ to proceed, we have to define the resonant/nonresonant decomposition of the phase-space.

\begin{defn}
For any $n>0$, we denote by $$\mathcal{M}_{2n}:=\left\{k=\left\{k_j\right\} \in \mathbb{Z}^{2n} \mbox{ s.t. } \sum_{j=1}^n k_j= \sum_{j=n+1}^{2n} k_j, \; \sum_{j=1}^n k^2_j= \sum_{j=n+1}^{2n} k^2_j\right\}$$
\end{defn}
Write $$Z_6=\sum_{\substack{k\in \mathcal{M}_6}}\tilde{Z}_{6,k}\psi_{k_1}\psi_{k_2}\psi_{k_3}\bar{\psi}_{k_4}\bar{\psi}_{k_5}\bar{\psi}_{k_6},$$
computing $$\cR_6=\left\{\Phi_{{\tt k},2},Z_6\right\},$$
one gets \begin{equation}\label{R6}\cR_6=\sum_{\substack{k\in \mathcal{M}_6 }} Z_{6,k,\tt k}\end{equation}
with
$$Z_{6,k,\tt k}:=-i\tilde{Z}_{6,k}\left(\delta_{k_1,\tt k}+\delta_{k_2,\tt k}+\delta_{k_3,\tt k}-\delta_{k_4,\tt k}-\delta_{k_5,\tt k}-\delta_{k_6,\tt k}\right)\psi_{k_1}\psi_{k_2}\psi_{k_3}\bar{\psi}_{k_4}\bar{\psi}_{k_5}\bar{\psi}_{k_6},$$
where   $\delta_{j,\tt k}$ is Kronecker's delta.

We introduce a function $\rho\in \mathcal{C}^{\infty}_0$, s.t. 
\begin{equation}
\rho(x)=\left\{\begin{array}{ll}
1 \mbox{ if } |x|>2\\0 \mbox{ if } |x|<1
\end{array}
\right..
\end{equation}
Recalling that $\omega_j:=c_2\left(|\psi_j|^2+\sum_k|\psi_k|^2\right)$, we denote by \begin{align}\nonumber a_k(\psi):=&\frac{1}{c_2}\left(\omega_{k_1}+\omega_{k_2}+\omega_{k_3}-\omega_{k_4}-\omega_{k_5}-\omega_{k_6}\right)\\ =&(|\psi_{k_1}|^2+|\psi_{k_2}|^2+|\psi_{k_3}|^2-|{\psi}_{k_4}|^2-|{\psi}_{k_5}|^2-|{\psi}_{k_6}|^2)
\end{align}
and, given $0<\delta<1$, we define the decomposition
$\cR_6:=\cR_6^{NR}+\cR_6^{R}$ with 
$$
\cR_6^{NR}  :=\sum_kZ_{6,k,\tt k}\rho\left(\frac{a_k(\psi)}{\delta}\right)$$  and 
$$\cR_6^{R}:=\sum_kZ_{6,k,\tt k}\left(1-\rho\left(\frac{a_k(\psi)}
   {\delta}\right)\right). $$
We define $\tilde \Phi_{\tk,6}$ to be the solution of equation
\eqref{nonres}, which is explicitely given by
$$\tilde \Phi_{\tk,6}:=i\sum_{\substack{k\in \mathcal{M}_6 }}\frac{Z_{6,k,\tt k}}{c_2a_k(\psi)}\rho\left(\frac{a_k(\psi)}{\delta}\right).$$  
\begin{oss}
$\tilde \Phi_{\tk,6}(\psi)\in P^2\left(6,\left\{\frac{A_i}{\delta^i}\right\}_{i=0}^2\right)\subset P^2\left(6,\left\{\frac{A}{\delta^i}\right\}_{i=0}^2\right)$ with $A:=\max_i A_i$.
\end{oss}

Finally we define the approximate integral of motion is given by
\begin{equation}\label{integraleapprossimato}
{\Phi}_{\tt k}^{(6)}:=\Phi_{{\tt k},2}+\Phi_{{\tt
    k},4}+\Phi_{{\tt k},6}+\tilde
\Phi_{\tk,6}+L_{\chi_4}\tilde \Phi_{\tk,6}. 
\end{equation}

The following lemma gives the structure of its time derivative. 
\begin{lem}\label{derivative}
Write \begin{align*}\left\{H,{\Phi}_{\tt k}^{(6)}\right\}=-\cR_6^{R}+R 
\end{align*}
then
\begin{equation}\label{erre}
R=\sum_{j=4}^{q+1}R_{2j}+\sum_{j=5}^{q+2}R_{2j,1}+\sum_{j=6}^{q+3}R_{2j,2}+\sum_{j=7}^{q+5}R_{2j,3},
\end{equation}
with $R_{2j}\in P_{2j},$
and there exists $C>0$ s.t. 
$$R_{2j,l}\in P^{3-l}\left(2j,\left\{\frac{C}{\delta^{m+l}}\right\}_{m=0}^{3-l}\right).$$
\end{lem}
\proof
One has
\begin{align}
\nonumber&\left\{H,{\Phi}_{\tt k}^{(6)}\right\}=\left\{H_2,\Phi_{\tk,2}\right\}\\
\label{ordine 4}&+\left\{H_2,\Phi_{\tk,4}\right\}+\left\{H_4,\Phi_{\tk,2}\right\}+\left\{H_2,\tilde\Phi_{\tk,6}\right\}\\
\label{ordine 6}&+\left\{Z_6,\Phi_{k,2}\right\}+\left\{Z_4,\tilde
\Phi_{\tk,6}\right\}+\left\{H_4^{R_{H_2}},\tilde
\Phi_{\tk,6}\right\}+\left\{H_2,L_{\chi_4}\tilde
\Phi_{\tk,6}\right\}\\
\label{ordine 8 e più}&+\sum_{j=2}^{n-2}\left(\left\{H_{2j},\Phi_{k,6}\right\}+\left\{H_{2j},L_{\chi_4}\tilde
\Phi_{\tk,6}\right\}+\left\{H_{2(j+1)},\Phi_{k,4}\right\}+\left\{H_{2(j+1)},\tilde
\Phi_{\tk,6}\right\}+\left\{H_{2(j+2)},\Phi_{k,2}\right\}\right)\\
\label{ordine 2n+2}&+\left\{H_{2(n-1)},\Phi_{k,6}\right\}+\left\{H_{2(n-1)},L_{\chi_4}\tilde
\Phi_{\tk,6}\right\}+\left\{H_{2n},\Phi_{k,6}\right\}+\left\{H_{2n},\tilde
\Phi_{\tk,6}\right\}\\
\label{ordine 2n+4}&+\left\{H_{2n},\Phi_{k,6}\right\}+\left\{H_{2n},L_{\chi_4}\tilde
\Phi_{\tk,6}\right\}.
\end{align}
Due to the construction, we have that $\left\{H_2,\Phi_{\tk,2}\right\}=0$ and $\left\{H_2,\Phi_{\tk,4}\right\}=-\left\{H_4,\tilde\Phi_{\tk,2}\right\}$. Due to the fact that $a_k$ and $\rho$ depend on the actions only and $\left\{Z_{6,k,\tt k},H_2\right\}=0$, one has  $\left\{H_2,\tilde
\Phi_{\tk,6}\right\}=0$ so that \eqref{ordine 4} vanishes.\\
Since $Z_4$ is a function of the actions only, we have also
 $$\left\{Z_4,\tilde
\Phi_{\tk,6}\right\}=i\sum_k\left\{Z_4,Z_{6,k,\tt k}\right\}\frac{\rho\left(\frac{a_k(\psi)}{\delta}\right)}{c_2a_k(\psi)}=\sum_k Z_{6,k,\tt k}\rho\left(\frac{a_k(\psi)}{\delta}\right)=\cR_6^{NR}.
$$
We note that $\left\{H_4^{R_{H_2}},\tilde
\Phi_{\tk,6}\right\}=-\left\{H_2,L_{\chi_4}\tilde
\Phi_{\tk,6}\right\}$ in fact, by the definition of $\chi_4$ and $\left\{H_2,\tilde
\Phi_{\tk,6}\right\}=0$, one has 
$$\left\{H_2,L_{\chi_4}\tilde
\Phi_{\tk,6}\right\}=-\left\{H_2,\left\{L^{-1}_{H_2}H_4^{R_{H_2}},\tilde
\Phi_{\tk,6}\right\}\right\}=$$
$$=\left\{L^{-1}_{H_2}H_4^{R_{H_2}},\left\{\tilde
\Phi_{\tk,6},H_2\right\}\right\}+\left\{\tilde
\Phi_{\tk,6},L_{H_2}L^{-1}_{H_2}H_4^{R_{H_2}}\right\}=\left\{\tilde
\Phi_{\tk,6},H_4^{R_{H_2}}\right\}.$$
So, by \eqref{R6},  line \eqref{ordine 6} reduces to $\sum_k Z_{6,k,\tt k}\left(\rho\left(\frac{a_k(\psi)}{\delta}\right)-1 \right)=-\cR_6^{R} $.\\
It remains to study now \eqref{ordine 8 e più}, \eqref{ordine 2n+2}and \eqref{ordine 2n+4}.
Using Lemma $\ref{parentesi_poisson_lemma}$, we have $$\left\{H_{2j},\tilde
\Phi_{\tk,6}\right\}=F_{1,j}+F_{2,j},$$
$$F_{1,j}\in P^2\left(2j+4,\left\{\frac{C}{\delta^{i+1}}\right\}_{i=0}^2\right),\; F_{2,j}\in P^1\left(2j+6,\left\{\frac{C}{\delta^{i+2}}\right\}_{i=0}^1\right),$$
$$L_{\chi_4}\tilde
\Phi_{\tk,6}=E_1+E_2, \; E_1\in P^1\left(8,\left\{\frac{C}{\delta^{i+1}}\right\}_{i=0}^2\right),\; E_2\in P^2\left(10,\left\{\frac{C}{\delta^{i+2}}\right\}_{i=0}^1\right),$$
so  $$\left\{H_{2j},L_{\chi_4}\tilde
\Phi_{\tk,6}\right\}=F_{3,j}+F_{4,j}+F_{5,j}, $$
$$F_{3,j}\in P^2\left(2j+6,\left\{\frac{C}{\delta^{i+1}}\right\}_{i=0}^2\right),\;F_{4,j}\in P^1\left(2j+8,\left\{\frac{C}{\delta^{i+2}}\right\}_{i=0}^1\right),$$
$$F_{5,j}\in P^0\left(2j+10,\frac{C}{\delta^3}\right), $$
$$\left\{H_{2j},\Phi_{k,2}\right\}\in P_{2j},\; $$
$$\left\{H_{2j},\Phi_{k,4}\right\}\in P_{2j+2},$$
$$\left\{H_{2j},\Phi_{k,6}\right\}\in P_{2j+4}.$$
$\\$\qed

\section{Measure estimates}\label{stime}
In this section we estimate $\||\psi_{\tt k}|^2\|_{\mu_{\beta}}$, $\|{\Phi}^{(6)}_{\tt k}\|_{\mu_{\beta}}$ and $\left\|\left\{H,{\Phi}^{(6)}_{\tt k}\right\}\right\|_{\mu_{\beta}}$.\\
 \begin{lem}\label{stimaresto}
There exists a constant $C>0$ s.t. for any $\beta>0$, $\delta$ s.t. $0<\delta\beta<1$,
one has
\begin{equation}\label{differenza azioni e cost moto}\|\Phi^{(6)}_{\tt k}-|\psi_{\tt_k}|^2\|^2_{g,\beta}\leq \frac{C}{\left(1+{\tt k}^2\right)^2\delta^2\beta^6},\end{equation}
\begin{equation}\label{stima resto}\|R\|_{g,\beta}^2\leq \frac{C}{\left(1+{\tt k}^2\right)^2\delta^6\beta^{14}},\end{equation}
where $R$ is defined by \eqref{erre}.
\end{lem}
\proof
We recall that $$\Phi^{(6)}_{\tt k}-|\psi_{\tt_k}|^2=\Phi_{{\tt k},4}+\Phi_{{\tt k},6}+\tilde\Phi_{{\tt k},6}-L_{\chi_4}\tilde\Phi_{{\tt k},6}.$$
So, using   Lemma \ref{stima_par_poisson_1} and Lemma $\ref{gau}$ with $M=2$,  we obtain   $$\|\Phi^{(6)}_{\tt k}-|\psi_{\tt_k}|^2\|^2_{g,\beta}\leq \frac{C}{\left(1+{\tt k}^2\right)^2}\left(\frac{1}{\beta^4}+\frac{1}{\beta^6}+\frac{1}{\delta^2\beta^6}+\frac{1}{\delta^2\beta^8}+\frac{1}{\delta^4\beta^{10}}\right)\leq$$
$$\leq \frac{5C}{\left(1+{\tt k}^2\right)^2\delta^2\beta^6}$$
where we used $0<\delta\beta<1$.
Using  \eqref{erre},  Lemma $\ref{derivative}$, Lemma \ref{stima_par_poisson_1} and  Lemma $\ref{gau}$ with $M=4$, we get
$$\|R\|_{g,\beta}^2\leq \frac{C}{\left(1+{\tt k}^2\right)^2}\left(\sum_{j=4}^{n+1}\frac{1}{\beta^{2j}}+\sum_{j=5}^{n+2}\frac{1}{\delta^2\beta^{2j}}+\sum_{j=6}^{n+3}\frac{1}{\delta^4\beta^{2j}}+\sum_{j=7}^{n+5}\frac{1}{\delta^6\beta^{2j}}\right)$$
so 
$$\|R\|_{g,\beta}^2\leq \frac{C}{\left(1+{\tt k}^2\right)^2\delta^6\beta^{14}}.$$
\endproof
It remains to estimate the resonant part, namely $\left\|\cR_6^{R}\right\|^2_{g,\beta}.$
\begin{lem}\label{resonantpart}
There exists a constant $\tilde{C}>0$ s.t. for any $\beta>0$  and $\delta>0$ s.t. $0<\delta\beta<1$,
one has
\begin{equation}\label{resonant}\left\|\cR_6^{R}\right\|^2_g\leq \tilde{C}\frac{(\delta\beta)^{\frac{2}{3}}}{\beta^6\left(1+{\tt k}^2\right)^2}.\end{equation}
\end{lem} 
The very technical proof is deferred to Appendix $\ref{stimetecniche}$. We remark that the difficult part consists in showing the presence of $\left(1+{\tt k}^2\right)^2$ at the denominators.

Finally, we obtain the following
\begin{lem}\label{stimaderivatagauss}
There exists  a constant ${C}>0$ s.t. for any $\beta>0$, one has
$$\left\|\dot\Phi^{(6)}_{\tt k}\right\|_{g,\beta}=\left\|\left\{H,\Phi^{(6)}_{\tt k}\right\}\right\|_{g,\beta}\leq \frac{C}{\left(1+{\tt k}^2\right)\beta^{3+\frac{1}{10}}}.$$
\end{lem}
\proof
We can choose $\delta$ in such a way that \eqref{stima resto} and \eqref{resonant} have the same size:$$\frac{1}{\delta^6\beta^{14}}=\frac{(\delta\beta )^{\frac{2}{3}}}{\beta^6}.$$
It follows that $\delta=\frac{1}{\beta^{\frac{13}{10}}}$ and the thesis. 
\endproof

Finally, using these results and Lemma $\ref{gibbsgauss}$, we obtain
\begin{lem}\label{stimaderivatacompleta}
There exists $\beta^*,{C}>0$ s.t. for any $\beta>\beta^*$, one has
$$\left\|\dot\Phi^{(6)}_{\tt k}\right\|_{\mu_{\beta}}\leq  \frac{C}{\left(1+{\tt k}^2\right)\beta^{3+\frac{1}{10}}}.$$
\end{lem}
\proof
This results is a simple consequence of Lemma $\ref{stimaderivatagauss}$ and Lemma $\ref{gibbsgauss}$.

\endproof

\section{Proof of Theorem \ref{consaz}}\label{dimteo}
\textit{Proof of Theorem} $\ref{consaz}$
Using Chebyshev's inequality, one has
\begin{equation}
\mu_{\beta} \left\{\psi : |{\Phi}^{(6)}_{\tt k}(\psi(t))-{\Phi}^{(6)}_{\tt k}(\psi)|>\eta_1\||\psi_{\tt k}|^2\|_{\mu_{\beta}}\right\}\leq \frac{\left\|{\Phi}^{(6)}_{\tt k}(\psi(t))-{\Phi}^{(6)}_{\tt k}(\psi)\right\|_{\mu_{\beta}}^2}{\eta_1^2\||\psi_{\tt k}|^2\|_{\mu_{\beta}}^2}.
\end{equation}
But ${\Phi}^{(6)}_{\tt k}(\psi(t))-{\Phi}^{(6)}_{\tt k}(\psi)=\int_0^t\dot{\Phi}^{(6)}_{\tt k}(\psi(s))ds$, so 
$$\left\|{\Phi}^{(6)}_{\tt k}(\psi(t))-{\Phi}^{(6)}_{\tt k}(\psi)\right\|_{\mu_{\beta}}\leq\int_0^t\left\|\dot{\Phi}^{(6)}_{\tt k}(\psi(s))\right\|_{\mu_{\beta}}ds$$ and, thanks to the invariance of the measure, we have
$$\left\|{\Phi}^{(6)}_{\tt k}(\psi(t))-{\Phi}^{(6)}_{\tt k}(\psi)\right\|_{\mu_{\beta}}\leq t\left\|\dot{\Phi}^{(6)}_{\tt k}\right\|_{\mu_{\beta}}.$$
So,
\begin{equation}\label{consquasicostmoto}
\mu_{\beta} \left\{\psi : |{\Phi}^{(6)}_{\tt k}(\psi(t))-{\Phi}^{(6)}_{\tt k}(\psi)|>\eta_1\||\psi_{\tt k}|^2\|_{\mu_{\beta}}\right\}\leq t^2\frac{\left\|\dot{\Phi}^{(6)}_{\tt k}\right\|_{\mu_{\beta}}^2}{\eta_1^2\||\psi_{\tt k}|^2\|^2_{\mu_{\beta}}}\leq\eta_2
\end{equation}
for any $ |t|<\frac{\eta_1\sqrt{\eta_2}\beta^{2+\frac{1}{10}}}{C}$, where we used Lemmas \ref{stimaazione} and \ref{stimaderivatacompleta}.
Using this result, we can study the variation of the $\tt k$-action.
In fact
\begin{equation}
\mu_{\beta} \left\{\psi : \left||\psi_{\tt k}(t)|^2-|\psi_{\tt k}(0)|^2\right|>\eta_1\||\psi_{\tt k}|^2\|_{\mu_{\beta}}\right\}\leq 
\end{equation}
$$
\leq\mu_{\beta} \left\{\psi : \left|{\Phi}^{(6)}_{\tt k}(\psi(t))-{\Phi}^{(6)}_{\tt k}(\psi)\right|>\frac{\eta_1}{3}\||\psi_{\tt k}|^2\|_{\mu_{\beta}}\right\}$$
$$
+\mu_{\beta} \left\{\psi :\left|\Phi^{(6)}_{\tt k}-|\psi_{\tt_k}|^2\right|(t)>\frac{\eta_1}{3}\||\psi_{\tt k}|^2\|_{\mu_{\beta}}\right\}$$
$$+\mu_{\beta} \left\{\psi : \left|\Phi^{(6)}_{\tt k}-|\psi_{\tt_k}|^2\right|(0)>\frac{\eta_1}{3}\||\psi_{\tt k}|^2\|_{\mu_{\beta}}\right\}
$$
$$\leq \frac{\eta_2}{2}+18\frac{\left\|\Phi^{(6)}_{\tt k}-|\psi_{\tt_k}|^2\right\|^2_{\mu_{\beta}}}{\eta_1^2\left\||\psi_{\tt k}|^2\right\|^2_{\mu_{\beta}}}\leq \eta_2
$$
for any $0<\eta_1<\frac{C}{\eta_2\beta^{\frac{7}{10}}},\; |t|<\frac{\eta_1\sqrt{\eta_2}\beta^{2+\frac{1}{10}}}{C}$, where we used Chebyshev's inequality, the conservation of the Gibbs measure, \eqref{differenza azioni e cost moto} with $\delta=\frac{1}{\beta^{\frac{13}{10}}}$ and Lemma \ref{gibbsgauss} to estimate the second and the third term.
Then Theorem \ref{consaz} is obtained by reformulating this inequality.\\
\qed\\
\textit{Proof of Corollary $\ref{tutte}$}
We consider two sequences $\eta_{1,k}:=\eta_1(1+k^2)^{\frac12}$, $\eta_{2,k}:=\frac{\eta_2}{(1+k^2)}\sum_j\frac{1}{1+j^2}$.\\
For any $k\in \mathbb{Z}$ and any $\alpha<1/2$, we define
 $$\cI_{\alpha,k}:=\left\{\psi : \left||\psi_{k}(t)|^2-|\psi_{k}(0)|^2\right|\leq\frac{\eta_1}{(1+k^2)^{\alpha}\beta}\right\}.$$
Using Theorem \ref{consaz}, one has
$$
\mu_{\beta}(\cI_{\alpha,k}^c)\leq\mu_{\beta} \left\{\psi : \left||\psi_{k}(t)|^2-|\psi_{k}(0)|^2\right|>\frac{\eta_1}{(1+k^2)^{\frac12}\beta}\right\}= 
$$
$$\mu_{\beta} \left\{\psi : \left||\psi_{k}(t)|^2-|\psi_{k}(0)|^2\right|>\frac{\eta_{1,k}}{(1+k^2)\beta}\right\}\leq \eta_{2,k}$$
for any $ |t|<C'\eta_1\sqrt{\eta_2}\beta^{2+\varsigma}$.\\
Denote $\cI_{\alpha}:=\cup_k\cI_{\alpha,k}$, one has that 
\begin{equation}\mu_{\beta}\left(\cI_{\alpha}^c\right)\leq\sum_k\mu_{\beta}\left(\cI_{\alpha,k}^c\right)\leq \eta_2.
\end{equation}
 
\qed

\appendix
\section{Lemmas on Gaussian and Gibbs  measure}\label{sezionemisura}
First, we recall that both  Gibbs and Gaussian  measures are constructed with a limit procedure starting from the "finite dimensional" measure which, in the Gaussian case,  is defined  by $$\mu_{\beta,g,N}:=\frac{e^{-\frac{\beta}{2}\|\Pi_N(\psi)\|^2_{H^1}}}{Z_{g,N}(\beta)}=\frac{e^{-\frac{\beta}{2}\sum_{|k|<N}\left(1+k^2\right)|\psi_k|^2}}{Z_{g,N}(\beta)},$$
$$Z_{g,N}(\beta):=\int_{\Pi_N(H^s)}e^{-\frac{\beta}{2}\sum_{|k|<N}\left(1+k^2\right)|\psi_k|^2}\prod_{|k|<N}d\psi_k d\bar{\psi}_k ,$$
where  $\Pi_N\left(\{\psi_k\}_{k\in\mathbb{Z}}\right):=\{\psi_k\}_{|k|<N}$.
(See \cite{Bourgain94}).

\begin{lem}\label{mistron} Let $N$ be an integer,  $1>\gamma>0$, 
Then   there exists $\tilde{C}(\gamma)>0$ s.t. for any $\beta>0$ one has   
$$\frac{\int_{\Pi_{N}(H^s)}\prod_{|k|<N}\chi_{\left\{|\psi_k|<\frac{1}{\left(1+k^2\right)^{\frac{\gamma}{2}}\sqrt{\beta}}\right\}} 
e^{-\frac{\beta}{2}\left(1+k^2\right)|\psi_k|^2}d\psi_kd\bar{\psi}_k}{Z_{g,N}(\beta)}\geq e^{-\tilde{C}(\gamma)}.$$
Moreover ${\tilde{C}}$ is independent on $N$.
\end{lem}
\proof
Using the independence of all the variables, one gets
$$\frac{\int_{\Pi_{N}(H^s)}\prod_{|k|<N}\chi_{\left\{|\psi_k|<\frac{1}{\left(1+k^2\right)^{\frac{\gamma}{2}}\sqrt{\beta}}\right\}} 
e^{-\frac{\beta}{2} \left(1+k^2\right)|\psi_k|^2}d\psi_kd\bar{\psi}_k}{Z_{g,N}(\beta)}=
$$
$$=\prod_{|k|<N}\frac{2\pi\int_{0}^{\infty}\chi_{\left\{\rho_k<\frac{1}{\left(1+k^2\right)^{\frac{\gamma}{2}}\sqrt{\beta}}\right\}} 
e^{-\frac{\beta}{2} \left(1+k^2\right)\rho_k^2}\rho_kd\rho_k}{2\pi\int_{0}^{\infty}
e^{-\frac{\beta}{2}\left(1+k^2\right)\rho_k^2}\rho_kd\rho_k}=\prod_{|k|<N}\frac{\int_{0}^{\frac{ \left(1+k^2\right)^{1-\gamma} }{2}} 
e^{-z_k}dz_k}{\int_{0}^{\infty}
e^{-z_k}dz_k}=
$$
$$
=\prod_{|k|<N}\left(1-e^{-\frac{\left(1+k^2\right)^{1-\gamma} }{2}}\right)\geq \prod_{k\in \mathbb{Z}} \left(1-e^{-\frac{\left(1+k^2\right)^{1-\gamma} }{2}}\right)
$$
$$
=e^{\sum_{|k|\in\mathbb{Z}}\log\left(1-e^{-\frac{\left(1+k^2\right)^{1-\gamma} }{2}}\right)}=e^{-\tilde{C}(\gamma)}.
$$
\endproof
As $N\rightarrow \infty$,  we get the following lemma
\begin{lem}\label{conv}
Let $\gamma$ be $1>\gamma>0$. 
Then, for any $\beta>0$,  one has   
$$\lim_{N\rightarrow \infty}\frac{\int_{\Pi_{N}(H^s)}\prod_{|k|<N}\chi_{\left\{|\psi_k|<\frac{1}{\left(1+k^2\right)^{\frac{\gamma}{2}}\sqrt{\beta}}\right\}} 
e^{-\frac{\beta}{2}\left(1+ k^2\right)|\psi_k|^2}d\psi_kd\bar{\psi}_k}{Z_{g,N}(\beta)}$$
$$=\int_{H^s}\left(\prod^{\infty}_{k\in \mathbb{Z}}\chi_{\left\{|\psi_k|<\frac{1}{\left(1+k^2\right)^{\frac{\gamma}{2}}\sqrt{\beta}}\right\}} \right)
d\mu_{g,\beta}.$$
\end{lem}
\proof
For any $M>N$, $M\in \mathbb{N}$, one has 
\begin{align}\nonumber\int_{\Pi_{N}(H^s)}\prod_{|k|<N}\chi_{\left\{|\psi_k|<\frac{1}{\left(1+k^2\right)^{\frac{\gamma}{2}} \sqrt{\beta}}\right\}}\frac{ 
e^{-\frac{\beta}{2} \sum_{|k|<N}\left(1+k^2\right)|\psi_k|^2}\prod_{|k|<N}d\psi_kd\bar{\psi}_k}{Z_{g,N}(\beta)}\\\nonumber=\int_{\Pi_{M}(H^s)}\prod_{|k|<N}\chi_{\left\{|\psi_k|<\frac{1}{\left(1+k^2\right)^{\frac{\gamma}{2}} \sqrt{\beta}}\right\}}\frac{ 
e^{-\frac{\beta}{2} \sum_{|k|<M}\left(1+k^2\right)|\psi_k|^2}\prod_{|k|<M}d\psi_kd\bar{\psi}_k}{Z_{g,M}(\beta)} 
\end{align}
So, one has 
$$\lim_{\substack{M\rightarrow\infty}}\int_{\Pi_{M}(H^s)}\prod_{|k|<N}\chi_{\left\{|\psi_k|<\frac{1}{\left(1+k^2\right)^{\frac{\gamma}{2}} \sqrt{\beta}}\right\}}\frac{ 
e^{-\frac{\beta}{2} \sum_{|k|<M}\left(1+k^2\right)|\psi_k|^2}\prod_{|k|<M}d\psi_kd\bar{\psi}_k}{Z_{g,M}(\beta)}=$$
$$=\int_{H^s}\prod_{|k|<N}\chi_{\left\{|\psi_k|<\frac{1}{\left(1+k^2\right)^{\frac{\gamma}{2}} \sqrt{\beta}}\right\}} 
d\mu_{g,\beta}.$$
But $\prod_{|k|<N}\chi_{\left\{|\psi_k|<\frac{1}{\left(1+k^2\right)^{\frac{\gamma}{2}} \sqrt{\beta}}\right\}}\rightarrow\prod_{k\in\mathbb{Z}}\chi_{\left\{|\psi_k|<\frac{1}{\left(1+k^2\right)^{\frac{\gamma}{2}} \sqrt{\beta}}\right\}}$ a.e. on $H^s$ as $N\rightarrow \infty$. Since $1\in  L^1{(H^s,\mu_{g,\beta})}$ and $\prod_{|k|<N}\chi_{\left\{|\psi_k|<\frac{1}{\left(1+k^2\right)^{\frac{\gamma}{2}} \sqrt{\beta}}\right\}}\leq 1$, by Lebesgue's dominated convergence Theorem, 
$$\lim_{N\rightarrow \infty}\int_{H^s}\prod_{|k|<N}\chi_{\left\{|\psi_k|<\frac{1}{\left(1+k^2\right)^{\frac{\gamma}{2}} \sqrt{\beta}}\right\}} 
d\mu_{g,\beta}=\int_{H^s}\lim_{N\rightarrow \infty}\prod_{|k|<N}\chi_{\left\{|\psi_k|<\frac{1}{\left(1+k^2\right)^{\frac{\gamma}{2}} \sqrt{\beta}}\right\}} 
d\mu_{g,\beta}=$$
$$=\int_{H^s}\prod_{k\in\mathbb{Z}}\chi_{\left\{|\psi_k|<\frac{1}{\left(1+k^2\right)^{\frac{\gamma}{2}} \sqrt{\beta}}\right\}} 
d\mu_{g,\beta}.$$

\endproof

\begin{oss}\label{misinf}
From Lemma $\ref{mistron}$ and Lemma $\ref{conv}$, we know that, if $1>\gamma> 0$ and $\beta>0$,    one has  
\begin{equation}\int_{H^s}\prod_{k\in\mathbb{Z}}\chi_{\left\{|\psi_k|<\frac{1}{\left(1+k^2\right)^{\frac{\gamma}{2}} \sqrt{\beta}}\right\}} 
d\mu_{g,\beta}\geq e^{-\tilde{C}(\gamma)}.
\end{equation}
\end{oss}
\begin{lem}\label{misura}
There exists a constant $\tilde{C}>0$ and $\beta^*>0$  s.t., for any
$\beta>\beta^*$,  one has
\begin{equation} 
1\geq\int_{H^s} e^{-\beta P}d\mu_{g,\beta}\geq e^{-2\tilde{C}}.
\end{equation}
\end{lem}
\proof
We remark that $P=\sum_{j=2}^qH_{2j}= \sum_{j=2}^q\frac{c_j}{2j}\|\psi\|^{2j}_{L^{2j}}$.\\
The first inequality is obvious.\\
We analyze now the second inequality.
By the definition of $P$, if we fix $s_1$, by Sobolev's inequality $H^{s_1}(\mathbb{T})\subset L^r(\mathbb{T}) \mbox{ if } r\in[1,\frac{2}{1-2s_1}]$. Therefore, choosing $\frac{q-1}{2q}<s_1<\frac{1}{2}$, there exists a costant $C_{sob}$  s.t.\begin{equation}\label{sobolev}\|\psi\|_{L^{2j}}<C_{sob}^{\frac{1}{2j}}\|\psi\|_{H^{s_1}}, \; j=2,...,q.\end{equation}
We fix $\frac{1}{2}+s_1<\gamma<1$, denote $D':=\sum_{j\in \mathbb{Z}}\frac{1}{\left(1+j^2\right)^{\gamma-s_1}}$, then we have:
$$\int_{H^s} e^{-\beta P}d\mu_{g,\beta}\geq \int_{H^{s}}\chi_{\left\{\|\psi\|^2_{H^{s_1}}\leq \frac{D'}{\beta}\right\}}e^{-\beta P}d\mu_{g,\beta}\geq$$
$$\int_{H^{s}}\chi_{\left\{\|\psi\|^2_{H^{s_1}}\leq \frac{D'}{\beta}\right\}}e^{-\frac{C_{sob}}{\beta}\left( \sum_{\substack{j=2,...,q\\c_j\geq0}}\frac{c_jD'^j}{\beta^{j-1}}\right)}d\mu_{g,\beta}\geq \int_{H^{s}}\chi_{\left\{\|\psi\|^2_{H^{s_1}}\leq \frac{D'}{\beta}\right\}}e^{-\frac{C_{sob}}{\beta}q\max_j{c_j}D'^j}d\mu_{g,\beta}$$
$$\geq e^{-\frac{C_{sob}}{\beta}q\max_j{c_j}D'^j}\int_{H^s}\prod_{k\in\mathbb{Z}}\chi_{\left\{|\psi_k|<\frac{1}{\left(1+k^2\right)^{\frac{\gamma}{2}} \sqrt{\beta}}\right\}} 
d\mu_{g,\beta}$$
$$\geq e^{-\frac{C_{sob}}{\beta}q\max{c_j}D'^j}e^{-\tilde{C}(\gamma)}\geq e^{-2\tilde{C}(\gamma)},$$
where the inequalities in the last line are true thanks to  Lemma $\ref{conv}$ and   for $\beta$ sufficiently large.
\endproof

\begin{oss}
$\mu_{\beta}$ is a good probability measure on $H^s$ since $\mu_{\beta}<\mu_{g,\beta}$ and $e^{-2\tilde{C}(\gamma)}\leq \frac{Z(\beta)}{Z_g(\beta)}\leq1$.
\end{oss}
For the proof is sufficient to note that 
$$\int_{H^s} e^{-\beta\left(
  \sum_{i=4}^n\frac{c_i}{i}\|\psi\|^i_{L^i}\right)}d\mu_{g,\beta}=\frac{Z(\beta)}{Z_g(\beta)}.$$

Using this result, we can obtain  Lemma $\ref{gibbsgauss}$ to estimate the $L^2$-norm in the Gibbs measure with the norm in Gaussian measure.
\\
\textit{Proof of Lemma $\ref{gibbsgauss}$ } 
We have $$\|f\|_{\mu_{\beta}}^2=\int_{H^s}|f|^2 d\mu_{\beta}\leq \frac{\int_{H^s}|f|^2d\mu_{g,\beta}}{\int_{H^s} e^{-\beta P}d\mu_{g,\beta}}
$$
and, from Lemma $\ref{misura}$, 
$$\|f\|_{\mu_{\beta}}^2\leq  \|f\|^2_{g,\beta} e^{2\tilde{C}(\gamma)}.
$$
\qed\\
\textit{Proof of Lemma $\ref{stimabasso}$}
As above we fix $\frac{q-1}{2q}<s_1<\frac{1}{2}$ and $\frac{1}{2}+s_1<\gamma<1$, we denote $D':=\sum_{j\in \mathbb{Z}}\frac{1}{\left(1+j^2\right)^{\gamma-s_1}}$, so we have:
$$\|f\|_{\mu_{\beta}}^2=\int_{H^s}|f|^2d\mu_{\beta}\geq \int_{H^s}|f|^2e^{-\beta P}d\mu_{g,\beta}\geq$$
$$\geq \int_{H^{s}}|f|^2\chi_{\left\{\|\psi\|^2_{H^{s_1}}\leq \frac{D'}{\beta}\right\}}e^{-\beta P}d\mu_{g,\beta}\geq$$
$$\geq e^{-\frac{C_{sob}}{\beta}q\max_j{c_j}D'^j}\int_{H^{s}}|f|^2\chi_{\left\{\|\psi\|^2_{H^{s_1}}\leq \frac{D'}{\beta}\right\}}d\mu_{g,\beta}$$
$$=e^{-\frac{C_{sob}}{\beta}q\max_j{c_j}D'^j}\left\|f\chi_{\left\{\|\psi\|^2_{H^{s_1}}\leq \frac{D'}{\beta}\right\}}\right\|_{g,\beta}^2.$$
\qed

We are now ready to give the proof of Lemma $\ref{stimaazione}$, namely the estimate from below of the $L^2$-norm of the actions in Gibbs measure.\\
\textit{Proof of Lemma $\ref{stimaazione}$}
We fix $\frac{q-1}{2q}<s_1<\frac{1}{2}$ and $\frac{1}{2}+s_1<\gamma<1$, we denote $D':=\sum_{j\in \mathbb{Z}}\frac{1}{\left(1+j^2\right)^{\gamma-s_1}}$, so
$$\left\||\psi_{\tt k}|^2\chi_{\left\{\|\psi\|^2_{H^{s_1}}\leq \frac{D}{\beta}\right\}}\right\|_{g,\beta}^2
\geq\int_{H^s}|\psi_{\tt k}|^4\prod_{k\in\mathbb{Z}}\chi_{\left\{|\psi_j|<\frac{1}{\left(1+j^2\right)^{\frac{\gamma}{2}} \sqrt{\beta}}\right\}} 
d\mu_{g,\beta}=$$
\begin{equation}\label{azione_finita}\lim_{N\rightarrow \infty}\frac{\int_{\Pi_N(H^s)}|\psi_{\tt k}|^4\prod_{j\in\mathbb{Z}}\chi_{\left\{|\psi_j|<\frac{1}{\left(1+j^2\right)^{\frac{\gamma}{2}} \sqrt{\beta}}\right\}} 
e^{-\frac{\beta}{2}\sum_{|j|<N}\left(1+j^2\right)|\psi_j|^2}\prod_{|j|<N}d\psi_jd\bar{\psi}_j}{\int_{\Pi_N(H^s)} 
e^{-\frac{\beta}{2}\sum_{j<N}\left(1+j^2\right)|\psi_j|^2}\prod_{|j|<N}d\psi_jd\bar{\psi}_j}.
\end{equation}
Using the independence of the variables, we have that \eqref{azione_finita} is equal to
\begin{align}\nonumber& \frac{\int_{\mathbb{C}}|\psi_{\tt k}|^4\chi_{\left\{|\psi_{\tt k}|<\frac{1}{\left(1+{\tt k}^2\right)^{\frac{\gamma}{2}} \sqrt{\beta}}\right\}} 
e^{-\frac{\beta}{2}\left(1+{\tt k}^2\right)|\psi_{\tt k}|^2}d\psi_{\tt k}d\bar{\psi}_{\tt k}}{\int_{\mathbb{C}} 
e^{-\frac{\beta}{2}\left(1+{\tt k}^2\right)|\psi_{\tt k}|^2}d\psi_{\tt k}d\bar{\psi}_{\tt k}}\times \\
\label{azione_finita1}
\times \lim_{N\rightarrow \infty}&
\frac{\int_{\Pi_{N-1}(H^s)}\prod_{\substack{j\in\mathbb{Z}\\j\neq\tt k}}\chi_{\left\{|\psi_j|<\frac{1}{\left(1+j^2\right)^{\frac{\gamma}{2}} \sqrt{\beta}}\right\}} 
e^{-\frac{\beta}{2}\sum_{\substack{|j|<N\\j\neq \tt k}}\left(1+j^2\right)|\psi_j|^2}\prod_{\substack{|j|<N\\j\neq \tt k}}d\psi_jd\bar{\psi}_j}{\int_{\Pi_{N-1}(H^s)} 
e^{-\frac{\beta}{2}\sum_{\substack{|j|<N\\j\neq \tt k}}\left(1+j^2\right)|\psi_j|^2}\prod_{\substack{|j|<N\\j\neq \tt k}}d\psi_jd\bar{\psi}_j},
\end{align}
Furthermore, since $$\frac{\int_{\mathbb{C}}\chi_{\left\{|\psi_{\tt k}|<\frac{1}{\left(1+{\tt k}^2\right)^{\frac{\gamma}{2}} \sqrt{\beta}}\right\}} 
e^{-\frac{\beta}{2}\left(1+{\tt k}^2\right)|\psi_{\tt k}|^2}d\psi_{\tt k}d\bar{\psi}_{\tt k}}{\int_{\mathbb{C}} 
e^{-\frac{\beta}{2}\left(1+{\tt k}^2\right)|\psi_{\tt k}|^2}d\psi_{\tt k}d\bar{\psi}_{\tt k}}<1,$$
one has that \eqref{azione_finita1} is lower than
\begin{align}\nonumber& \frac{\int_{\mathbb{C}}|\psi_{\tt k}|^4\chi_{\left\{|\psi_{\tt k}|<\frac{1}{\left(1+{\tt k}^2\right)^{\frac{\gamma}{2}} \sqrt{\beta}}\right\}} 
e^{-\frac{\beta}{2}\left(1+{\tt k}^2\right)|\psi_{\tt k}|^2}d\psi_{\tt k}d\bar{\psi}_{\tt k}}{\int_{\mathbb{C}} 
e^{-\frac{\beta}{2}\left(1+{\tt k}^2\right)|\psi_{\tt k}|^2}d\psi_{\tt k}d\bar{\psi}_{\tt k}}\times \\
\nonumber
\times \lim_{N\rightarrow \infty}&
\frac{\int_{\Pi_{N}(H^s)}\prod_{\substack{j\in\mathbb{Z}}}\chi_{\left\{|\psi_j|<\frac{1}{\left(1+j^2\right)^{\frac{\gamma}{2}} \sqrt{\beta}}\right\}} 
e^{-\frac{\beta}{2}\sum_{\substack{|j|<N}}\left(1+j^2\right)|\psi_j|^2}\prod_{\substack{|j|<N}}d\psi_jd\bar{\psi}_j}{\int_{\Pi_{N}(H^s)} 
e^{-\frac{\beta}{2}\sum_{\substack{|j|<N}}\left(1+j^2\right)|\psi_j|^2}\prod_{\substack{|j|<N}}d\psi_jd\bar{\psi}_j}
\end{align}
$$\geq \frac{\int_0^{\frac{1}{\left(1+{\tt k}^2\right)^{\frac{\gamma}{2}}\sqrt{\beta}}}\rho_{\tt k}^5 
e^{-\frac{\beta}{2}\left(1+{\tt k}^2\right)\rho_{\tt k}^2}d\rho_{\tt k}}{\int_0^{\infty}\rho_{\tt k}
e^{-\frac{\beta}{2}\left(1+{\tt k}^2\right)\rho_{\tt k}^2}d\rho_{\tt k}}\int_{H^s}\prod_{k\in\mathbb{Z}}\chi_{\left\{|\psi_j|<\frac{1}{\left(1+j^2\right)^{\frac{\gamma}{2}} \sqrt{\beta}}\right\}} 
d\mu_{g,\beta}
$$
$$\geq  \frac{4}{\beta^2\left(1+{\tt k}^2\right)^2}\int_0^{\frac{\left(1+{\tt k}^2\right)^{1-\gamma}}{2}}z_{\tt k}^2e^{-z_{\tt k}} 
dz_k  e^{-2\tilde{C}(\gamma)}$$
$$\geq  \frac{ e^{-\tilde{C}(\gamma)}}{\beta^2\left(1+{\tt k}^2\right)^2}\int_0^{\frac{\left(1+{\tt k}^2\right)^{1-\gamma}}{2}}z_{\tt k}^2e^{-z_{\tt k}} 
dz_{\tt k}\geq  \frac{ e^{-\tilde{C}(\gamma)}}{\beta^2\left(1+{\tt k}^2\right)^2}\int_0^{\frac{1}{2}}x^2e^{-x} 
dx,$$
where in the last line we use Lemma $\ref{conv}$. 
So, for  $\beta$ large enough, using  Lemma $\ref{stimabasso}$, one has
$$\||\psi_{\tt k}|^2\|_{\mu_{\beta}}^2\geq e^{-\frac{C_{sob}}{\beta}q\max_j{c_j}D'^q}\left\||\psi_{\tt k}|^2\chi_{\left\{\|\psi\|^2_{H^{s_1}}\leq \frac{D'}{\beta}\right\}}\right\|_{g,\beta}^2$$
$$\geq e^{-\frac{C_{sob}}{\beta}q\max_j{c_j}D'^q} \frac{ e^{-\tilde{C}(\gamma)}}{\beta^2\left(1+{\tt k}^2\right)^2}\int_0^{\frac{1}{2}}x^2e^{-x} 
dz_{\tt k}=\frac{C_1^2(\gamma)}{\beta^2\left(1+{\tt k}^2\right)^2}.  
$$ 
\qed

The support of the Gaussian measure is described in the following lemma in which the main part is that we specify the dependence on $\beta$ of the r.h.s.
\begin{lem}\label{grandideviazioni}
  For any $s_1<\frac{1}{2}$, $a<\frac{1}{2}$, $M>0$ and $\beta$ large
  enough, there exists a constant $C>0$ s.t.
$$\mu_{\beta}\left(\left\{\|\psi\|_{H^{s_1}}>M\right\}\right)\leq Ce^{-a\beta M^2}$$ 
\end{lem}
\proof
We consider 
$$e^{a\beta M^2}\mu_{\beta}\left(\left\{\|\psi\|_{H^{s_1}}>M\right\}\right)\leq e^{2\tilde{C}}e^{a\beta M^2}\mu_{g,\beta}\left(\left\{\|\psi\|_{H^{s_1}}>M\right\}\right)$$
$$= e^{2\tilde{C}}\int_{\{\|\psi\|_{H^{s_1}}>M\}\cap H^s}e^{a\beta M^2}d\mu_{g,\beta}\leq e^{2\tilde{C}}\int_{\{\|\psi\|_{H^{s_1}}>M\}\cap H^s}e^{a\|\psi\|^2_{H^{s_1}}}d\mu_{g,\beta}$$
$$\leq e^{2\tilde{C}}\int_{{H^{s}}}e^{a\beta \|\psi\|^2_{H^{s_1}}}d\mu_{g,\beta}=e^{2\tilde{C}}\int_{{H^{s}}}e^{a\beta \sum_j \left(1+j^2\right)^{s_1}|\psi_j|^2}d\mu_{g,\beta}$$
$$=e^{2\tilde{C}}\frac{\int_{{H^{s}}}e^{a\beta\sum_j\left(1+j^2\right)^{s_1}|\psi_j|^2-\frac{\beta}{2}\sum_j \left(1+j^2\right)|\psi_j|^2}\prod_j d\psi_jd\bar{\psi}_j}{\int_{{H^{s}}}e^{-\frac{\beta}{2}\sum_j \left(1+j^2\right)|\psi_j|^2}\prod_j d\psi_jd\bar{\psi}_j}$$
\begin{equation}\label{grandideviazioni}=e^{2\tilde{C}}\prod_j\frac{\int_{\mathbb{C}}e^{a\beta \left(1+j^2\right)^{s_1}|\psi_j|^2-\frac{\beta}{2} \left(1+j^2\right)|\psi_j|^2} d\psi_jd\bar{\psi}_j}{\int_{\mathbb{C}}e^{-\frac{\beta}{2}\left(1+j^2\right)|\psi_j|^2} d\psi_jd\bar{\psi}_j}\end{equation}
Using the substitution $\psi_j={\frac{\sqrt{2z_j}}{\sqrt{\beta(1+j^2)}}}e^{i\theta_j}$, $z_j\in \mathbb{R}^+$, $\theta_j\in [0,2\pi)$  and the fact  that $\int_{\mathbb{R}^+}e^{-z}dz=1$, one has that \eqref{grandideviazioni} is equal to
$$e^{2\tilde{C}}\prod_j \int_{0}^{\infty}e^{-\left(1-2a\left(1+j^2\right)^{s_1-1}\right)z_k}dz_k
$$
$$=e^{2\tilde{C}}\prod_j\left(1+\frac{2a}{\left(1+j^2\right)^{1-s_1}-2a}\right)=C.
$$
\endproof
\begin{oss}From the previous lemma, if $M$ goes to $+\infty$, we obtain that for any $s_1<\frac{1}{2}$, 
\end{oss}$$\mu_{\beta}\left(\left\{\|\psi\|_{H^{s_1}}=+\infty\right\}\right)=0.$$
In particular, we obtain that, for any $s_1>s$, $\mu_{\beta}\left(H^{s}\setminus H^{s_1}\right)=0$.\\
\textit{Proof of Lemma $\ref{momenti gaussiana}$}
Having fixed $\beta$ large enough, $n>0$, and $a<\frac{\beta}{2}$, there exists  a constant $C>0$ s.t. for any $x>C$, $x^n<e^{ax^2}$, so,   one has
$$ \int_{H^{s}}\|\psi\|^n_{H^{s_1}}d\mu_{g,\beta}<\int_{\{\|\psi\|_{H^{s_1}}<C\}\cap H^s}\|\psi\|^n_{H^{s_1}}d\mu_{g,\beta} +\int_{\{\|\psi\|_{H^{s_1}}>C\}\cap H^s}e^{a\|\psi\|^2_{H^{s_1}}}d\mu_{g,\beta}$$
$$\leq C^n +\int_{H^s}e^{a\|\psi\|^2_{H^{s_1}}}d\mu_{g,\beta}=C^n+\prod_j\left(1+\frac{2a}{\beta\left(1+j^2\right)^{1-s_1}-2a}\right)<\infty,$$
where in the last line we proceed as in Lemma $\ref{grandideviazioni}$.
So we proved that $\|\psi\|^n_{H^{s_1}}\in L^1(H^{s},d\mu_{g,\beta})$.
By Lemma $\ref{gibbsgauss}$ we have  that $\|\psi\|^n_{H^{s_1}}\in L^1(H^{s},d\mu_{\beta})$.
\qed

\section{Technical lemmas}\label{stimetecniche}
\subsection{Proof of Lemma \ref{gau}}\label{Gaussian estimates}

We recall that, given a set $K$ of indices $(k_1,...,k_{2n})$ with an even number of
components, we denote
$$
K_1:=\{k_1,...,k_n  \}\ ,\quad K_2:=\{k_{n+1},...,k_{2n}  \} \ . 
$$

\begin{lem}\label{2indip.}
Let $k\in \mathbb{Z}^{2n}$ and $j\in \mathbb{Z}^{2m}$ be 2 integer vectors,
each one fulfilling the zero momentum condition and an $(M,\tk)$
admissible condition.

Assume that $K_1\cup J_2=K_2\cup J_1$, then there exist $x,y\in
K_1\cup J_2$ and a constant $C$, s.t. $|x|,|y|\geq
|\tk|/C$. Furthermore $\{x,y\}$ is uniquely determines by $K_1\cup
J_2\setminus \{x,y\}$.
\end{lem}
\proof For future reference we write the $(M,\tk)$
admissible conditions for the two vectors:
\begin{align}
  \label{ad.1}
  \sum_{i=1}^{2n}a_ik_i=\tk\ ,
  \\
  \label{ad.2}
  \sum_{i=1}^{2n}b_ij_i=\tk\ .
\end{align}
We give now a recoursive procedure in order to determine the elemnts $x,y$
in the statement. 

From \eqref{ad.1} there exists $l_1$ s.t. $\left|k_{l_1}\right|\geq
|\tk|/2nM$. By possibily interchanging $K_1\cup J_2$ with $K_2\cup
J_1$ and reordering the indexes, we can always assume that $l_1=1$. So
we have
$$
\left|k_1\right|\geq
\frac{|\tk|}{2nM}\ ,\quad a_1\not=0\ .
$$
In the following we will make several cases. 

We look for the ``companion'' of $k_1$ in $K_2\cup
J_1$. We have two possibilities:
\begin{itemize}
\item [(A)] It belongs to $J_1$ and therefore, by possibly reordering
  the indexes it is given by $j_1$ (thus we have $k_1=j_1$)
\item [(B)] It belongs to $K_2$ and therefore, by possibly reordering
  the indexes it is given by $k_{n+1}$ (thus we have $k_1=k_{n+1}$)
\end{itemize}

We begin by analyzing the case (A). We use the zero momentum condition
on $k$ in order to compute $k_1$ as a function of the other components
and we substitute in \eqref{ad.1}, which takes the form
\begin{equation}
  \label{ad.5}
\sum_{i=2}^{n}(a_i-a_1)k_i+\sum_{i=1}^n(a_{i+n}+a_1)k_{i+n}=\tk\ .
\end{equation}
Then there exists at least one of the $k_i$'s which has modulus larger
then a constant times $|\tk|$. There are two possibilities
\begin{itemize}
\item[(A.1)] It belongs to $K_1$, thus (up to reordering) it is given
  by $k_n$:
  \begin{equation}
    \label{ad.5.1}
|k_n|\geq \frac{|\tk|}{2(n-1)M}\quad \&\quad a_1\not=a_n
  \end{equation}
\item[(A.2)] It belongs to $K_2$, thus (up to reordering) it is given
  by $k_{2n}$:
  \begin{equation}
    \label{ad.5.2}
|k_{2n}|\geq \frac{|\tk|}{2(n-1)M}\quad \&\quad a_1\not=-a_{2n}\ .
  \end{equation}
\end{itemize}

We analyze first (A.1). Consider the companion of $k_n$, there are two
further possibilities:
\begin{itemize}
\item[(A.1.1)] It belongs to $J_1$, call it $j_m$ (thus $k_n=j_m$),
\item[(A.1.2)] It belongs to $K_2$, call it $k_{2n}$ (thus $k_n=k_{2n}$).
\end{itemize}

We analyze (A.1.1). In this case, given $K_1\cup
J_2\setminus\{k_1,k_n\}$ also $K_2\cup
J_1\setminus\{j_1,j_m\}$ is fixed. Then \eqref{ad.5} determines $k_n$
and then \eqref{ad.1} determines $k_1$. This concludes the case
(A.1.1).

We analyze now (A.1.2). Given $K_1\cup J_2\setminus\{k_1,k_n\}$ also
$K_2\cup J_1\setminus\{j_1,k_{2n}\}$ is fixed. So, also $J_1\cup
J_2\setminus \{j_1\}$ is determined. Then, by the zero momentum
condition on $j$ one determines $j_1=k_1$. Still one has to determine
$k_n=k_{2n}$. To this end one would like to use \eqref{ad.5}. This is
possible if the coefficients of $k_n$ and $k_{2n}$ do not cancel
out. If this happens, then consider
$k':=(k_1,...,k_{n-1},k_{n+1},...,k_{2n-1})$ and iterate the argument
of situation (A) with it (which also fulfills the zero momentum
condition). Iterating $n$ possibly decreases by one at each
step. Since $k'$ (and its iterates) has to fulfill an $(M,\tk)$
relation, which in particular is inhomogeneous, the procedure
terminates with a nontrivial $k'$ of dimension at least 2. This
concludes this case.

This concludes the analysis of (A.1).

We now analyze the case (A.2). We have two cases according to the
position of the companinon of $k_{2n}$.
\begin{itemize}
\item[(A.2.1)] It is $k_n\in K_{1}$ (thus $k_n=k_{2n}$)
\item[(A.2.2)] It is $j_{2m}\in J_{2}$ (thus $j_{2m}=k_{2n}$).
\end{itemize}
The situation of the case (A.2.1) is identical to that of (A.1.2) and
has already been analyzed. 

We study now (A.2.2). Given $K_1\cup J_2\setminus\{k_1,j_{2m}\}$ also
$K_1\cup K_2\setminus \{k_1,k_{2n}\}$ is determined. But, by the
second of \eqref{ad.5.2}, \eqref{ad.5} determines $k_{2n}$. Then $k_1$
is determined by \eqref{ad.1}.

{\bf This concludes the analysis of (A)}.

We come to (B). Substituting $k_1=k_{n+1}$ in \eqref{ad.1} we get
\begin{equation}
  \label{ad.7}
(a_1+a_{n+1})k_1+\sum_{i=2}^n(a_ik_i+a_{i+n}k_{i+n})=\tk\ .
\end{equation}
We have two possibilities
\begin{itemize}
\item[(B.1)] $-a_1\not =a_{n+1}$
  \item[(B.2)] $-a_1=a_{n+1}$
\end{itemize}

We analyze (B.1). We concentrate on $j$. By \eqref{ad.2} there exists
one of the $j_i$'s which is ``big''. There are two cases
\begin{itemize}
\item[(B.1.1)] it belongs to $J_1$ and thus it is $|j_1|\geq |\tk|/2mM$
\item[(B.1.2)] it belongs to $J_2$ and thus it is $|j_{2m}|\geq |\tk|/2mM$
\end{itemize}
Analyze (B.1.1). There are again two cases according to the companion
of $j_1$
\begin{itemize}
\item[(B.1.1.1)] It belongs to $K_1$, thus it is $k_n=j_1$.
\item[(B.1.1.2)] It belongs to $J_2$, thus it is $j_{m+1}=j_1$.
\end{itemize}
Analyze (B.1.1.1). Given $K_1\cup J_2\setminus\{k_1,k_n\}$ also
$K_2\cup J_1\setminus \{k_{n+1},j_{1}\}$ is determined. Thus also
$J_1\cup J_2\setminus\{j_1\}$ is determined. So, from the zero
momentum condition also $j_1=k_n$ is determined. From \eqref{ad.7}
also $k_1$ is determined.

We analyze (B.1.1.2). First we remark that given $K_1\cup
J_2\setminus\{k_1,j_{2n}\}$ also $K_2\cup J_1\setminus
\{k_{n+1},j_{n}\}$ is determined, thus $K_1\cup
K_2\setminus\{k_1,k_{n+1}\} $ is determined, and then, by \eqref{ad.7}
also $k_1=k_{n+1}$ is determined. Then we have to determine one
further large component.

Substituting $j_1=j_{m+1}$ in \eqref{ad.2} one gets
\begin{equation}
  \label{ad.8}
\sum_{i=2}^m(b_ij_i+b_{i+m}j_{i+m})+(b_1+b_{m+1})j_1=\tk  
\ .
\end{equation}
We have two cases
\begin{itemize}
\item[(B.1.1.2.1)] $b_{1}+b_{m+1}\not=0$
\item[(B.1.1.2.2)] $b_{1}+b_{m+1}=0$
\end{itemize}
Case (B.1.1.2.1). Given $K_1\cup J_2\setminus\{k_1,j_{m+1}\}$ also
$K_2\cup J_1\setminus \{k_{n+1},j_{1}\}$ is determined. Thus also
$J_1\cup J_2\setminus\{j_1,j_{m+1}\}$ is determined, but then one can
use \eqref{ad.8} to compute $j_{1}$. This concludes the analysis of this case.

Case (B.1.1.2.2). In this case \eqref{ad.8} becomes a $(2M,\tk)$
admissible condition for $j':=(j_2,...,j_{m},j_{m+2},...,j_{2m})$,
which also fulfills the zero momentum condition. Thus one is again in
the situation (B.1) but with $j'$ in place of $j$. Iterating the
construction one decreases $m$ at each step, and therefore the
procedure terminates in a finite number of steps.

We come to the case (B.1.2). We distinguish two cases according to
the position of the companion of $j_{2m}$.

\begin{itemize}
\item[(B.1.2.1)] It belongs to $K_2$, thus it is $k_{2n}$.
\item[(B.1.2.2)] It belongs to $J_1$, thus it is $j_{2m}$.
\end{itemize}

Case (B.1.2.1). Given $K_1\cup J_2\setminus\{k_1,j_{2m}\}$ also
$K_2\cup J_1\setminus \{k_{n+1},k_{2n}\}$ is determined. Thus also
$J_1\cup J_2\setminus\{j_{2m}\}$ is determined. Then by the zero
momentum condition on $j $ also $j_{2m}=k_{2n}$ is determined and one
can use \eqref{ad.7} to determine $k_1$.  

Case (B.1.2.2). By reasoning in a similar way one determines
$k_1=k_{n+1}$. Still one has to determine $j_{m}=j_{2m}$ and this can
be done exactly (up to a relabellin of the indexes) as in the case
(B.1.1.2). It means that if $b_{1}+b_{m+1}\not=0$ the argument is
complete, otherwise we have to start a recoursion as above in the case
(B.1.1.2.2).

In the case (B.2), \eqref{ad.7} becomes an $(M,\tk)$ admissible
condition for $k':=(k_2,...,k_n,k_{n+2},...,k_{2n})$ which also
fulfills the zero momentum condition. Thus  the construction is
repeated with $k'$ in place of $k$ and after a finite number of steps
the construction stops.

\endproof

We can now prove Lemma $\ref{gau}$.\\
\textit{Proof of Lemma $\ref{gau}$}
The proof is similar to that of Lemma $\ref{gausemplice}$. In the same
way, we get an estimate analogous to  \eqref{serie}, the only
difference is that the sum is not on $\mathcal{T}$ but on the set of
$(k,j)$ fulfilling the assumptions of Lemma \ref{2indip.}.
We denote  this set by $\tilde{\mathcal{T}}$.

So, we estimate 
\begin{equation}\label{serie_fuori_M}
\sum_{\substack{(k,j)\in \tilde{\mathcal{T}} }}\frac{1}{\prod_{i=1}^n\left(1+k^2_i\right)\left(1+j^2_{n+i}\right)}.
\end{equation}
If ${\tt k}=0$, then we can proceed exactly as in  Lemma $\ref{gausemplice}$.\\
If ${\tt k}\neq0$, we note that at most $[(2n!)]^2$
couples $(k,j)$ give the same set $K_{1}\cup J_{2}=K_2\cup J_1$.
So using Lemma \ref{2indip.}, we obtain 
\begin{align}
&\sum_{\substack{(k,j)\in \tilde{\mathcal{T}}
  }}\frac{1}{\prod_{i=1}^n\left(1+k^2_i\right)\left(1+j^2_{n+i}\right)}
\\ 
&\leq\frac{[(2n)!]^2}{\left(1+\left(\frac{{\tt
      k}}{C}\right)^2\right)^2}\sum_{l_1,...,l_{2n-2}}\frac{1}{\prod_{t=1}^{2m-2}(1+l_t^2)} 
\\
&\leq\frac{C}{\left(1+{\tt k}^2\right)^2}\left(\sum_{l}\frac{1}{(1+l^2)}\right)^{n-2}.
\end{align}
\qed

\subsection{Estimate of the resonant part}
First, we introduce a lemma useful to estimate the measure of the resonant region.\\
Given $n\in \mathbb{N}$ and  $k=(k_1,...,k_n)\in \mathbb{Z}^n$, we denote by $\tt M$ the cardinality of $Supp(k)$ and for any $\epsilon>0$, we define the non smooth cutoff function $$\chi(x)=
\left\{
\begin{array}{ll}
0 \mbox{ if } |x|\geq 1 \\
1 \mbox{ if } |x|
<1
\end{array}
\right.,\; \; \chi_{\epsilon}(x):=\chi\left(\frac{x}{\epsilon}\right).
$$

\begin{lem}\label{piccoli}
Let $0<\epsilon$, $n\in \mathbb{N}$, $k=(k_1,...,k_n)\in \mathbb{Z}^n$, $\{a_i\}_{i=1}^n\in \mathbb{Z}^n\setminus \{0\}$. Then there exists a constant $C(n)>0$  s.t.,
denoting $\tilde{k}:=\min_{l\in Supp(k), a_l\neq0}k_l$ and $\tilde{a}$ the correspondent coefficient in $\{a_i\}_{i=1}^n$,
\begin{equation}\label{trapiani}
\int_{\mathbb{R}_+^{\tt M}} \left(\prod_{i=1}^nz_{k_i}\right)
\chi\left(\sum_{i=1}^na_i\frac{z_{k_i}}{k_i^2}\right)e^{-\sum_{l\in Supp(k)} z_l}\prod_{l\in Supp(k)} dz_l\leq 4\tilde aC(n)\tilde{k}^2\epsilon.
\end{equation}

\end{lem}
\proof
We have that $z^le^{-z}<(2l)^le^{-l}e^{-\frac{z}{2}}<(2n)^ne^{-\frac{z}{2}}$, so, denoting by 
$I$ the left side of $\eqref{trapiani}$ and using the substitution $\frac{z_l}{2}=x_l$,
we have
 $$I\leq C_1(n)\int_{\mathbb{R}_+^{\tt M}}
\chi\left(\sum_{i=1}^n2a_i\frac{x_{k_i}}{k_i^2}\right)e^{-\sum_{l\in Supp(k)} x_l}\prod_{l\in Supp(k)} dx_l.$$
We denote $A(x):=\sum_{\substack{k_i\neq\tilde{k}}}2a_i\frac{x_{k_i}}{k_i^2}$.
So $I$ is bounded from above by
$$ C(n)\int_{\mathbb{R}_+^{{\tt M}-1}}\prod_{\substack{l\in Supp(k)\\l\neq\tilde k}} dx_le^{-\sum_{\substack{l\in Supp(k)\;l\neq\tilde{k}}} x_l}\int_{\left(-\epsilon-A(x)\right)\frac{\tilde{k}^2}{2\tilde{a}}}^{\left(\epsilon-A(x)\right)\frac{\tilde{k}^2}{2\tilde{a}}}e^{-x_{\tilde{k}}} dx_{\tilde k}$$
$$< C(n)\int_{\mathbb{R}_+^{{\tt M}-1}}\prod_{\substack{l\in Supp(k)\\l\neq\tilde k}}dx_le^{-\sum_{\substack{l\in Supp(k)\;l\neq\tilde{k}}} x_l}\int_{\left(-\epsilon-A(x)\right)\frac{\tilde{k}^2}{2\tilde{a}}}^{\left(\epsilon-A(x)\right)\frac{\tilde{k}^2}{2\tilde{a}}}dx_{\tilde k}=4\tilde aC(n)\tilde{k}^2\epsilon.$$
\endproof

\textit{Proof of Lemma} $\ref{resonantpart}$
 $$\left\|\cR_6^{R}\right\|^2_g=\left\|\sum_{\substack{k\in \mathcal{M}_6 }} Z_{6,k,\tt k}(\psi)\left(1-\rho\left(\frac{a_k(\psi)}{\delta}\right) \right)\right\|^2_g ,$$
so

$$\left\|\cR_6^{R}\right\|^2_g=$$
$$=\int_{H^s}\left(\sum_{\substack{k\in \mathcal{M}_6 }} Z_{6,k,\tt k}(\psi)\left(1-\rho\left(\frac{a_k(\psi)}{\delta}\right) \right)\right)\left(\sum_{\substack{j\in \mathcal{M}_6 }}\bar{Z}_{6,j,\tt k}(\psi)\left(1-\rho\left(\frac{a_j(\psi)}{\delta}\right) \right)\right)d\mu_{\beta}$$
\begin{equation}\label{serieintegrale}
=\int_{H^s}\sum_{\substack{k,j\in \mathcal{M}_6 }}Z_{6,k,\tt k}(\psi) \bar{Z}_{6,j,\tt k}(\psi)\left(1-\rho\left(\frac{a_j(\psi)}{\delta}\right) \right)\left(1-\rho\left(\frac{a_k(\psi)}{\delta}\right) \right)d\mu_{\beta}.
\end{equation}
As in Lemma $\ref{gau}$,  for Lemmas $\ref{convolution}$ and $\ref{momenti gaussiana}$, we can exchange the order between the integral and the series.\\
So $\eqref{serieintegrale}$ is equal to
$$\sum_{\substack{k,j\in \mathcal{M}_6 }}\int_{H^{s}} Z_{6,k,\tt k}(\psi) \bar{Z}_{6,j,\tt k}(\psi)\left(1-\rho\left(\frac{a_j(\psi)}{\delta}\right) \right)\left(1-\rho\left(\frac{a_k(\psi)}{\delta}\right) \right)d\mu_{\beta}.$$
We analyze, now one single term of the series, namely:
\begin{align}
&\tilde{Z}_{6,k}\left(\delta_{k_1,\tt k}+\delta_{k_2,\tt k}+\delta_{k_3,\tt k}-\delta_{k_4,\tt k}-\delta_{k_5,\tt k}-\delta_{k_6,\tt k}\right)\\
\times&\bar{\tilde{Z}}_{6,j}\left(\delta_{k_1,\tt k}+\delta_{k_2,\tt k}+\delta_{k_3,\tt k}-\delta_{k_4,\tt k}-\delta_{k_5,\tt k}-\delta_{k_6,\tt k}\right)\\ 
\times&\int\prod_{i=1}^3\psi_{j_i}\psi_{k_{3+i}}\bar{\psi}_{j_{3+i}}\bar{\psi}_{k_i}\left(1-\rho\left(\frac{a_j(\psi)}{\delta}\right) \right)\left(1-\rho\left(\frac{a_k(\psi)}{\delta}\right) \right)d\mu_{\beta}.
\end{align}
We remark that:
$$a_k(\psi):= (|\psi_{k_1}|^2+|\psi_{k_2}|^2+|\psi_{k_3}|^2-|{\psi}_{k_4}|^2-|{\psi}_{k_5}|^2-|{\psi}_{k_6}|^2).$$
With the transformation $\psi=r e^{i\theta}$, denoted by $S_{k,j}:=Supp(k,j)$,  the integral becomes 
$$\frac{\int_{r_k\in\mathbb{R}_+} \prod_{i=1}^6 r_{j_i}r_{k_i}\left(1-\rho\left(\frac{\tilde{a}_j(r)}{\delta}\right) \right)\left(1-\rho\left(\frac{\tilde{a}_k(r)}{\delta}\right) \right)e^{-\beta\sum_{l\in S_{k,j}}\left(1+ l^2\right)r_l^2}\prod_{k\in S_{k,j}} r_ldr_l}{\prod_{l\in S_{k,j}}\int_{\mathbb{R}_+}e^{-\beta\left(1+ l^2\right)r_l^2} l_kdr_l}$$
$$\times\frac{ \int_{\theta_k\in[0,2\pi]}e^{i(\theta_{j_1}+\theta_{j_2}+\theta_{j_3}+\theta_{k_4}+\theta_{k_5}+\theta_{k_6}-\theta_{j_4}-\theta_{j_5}-\theta_{j_6}-\theta_{k_1}-\theta_{k_2}-\theta_{k_3})}\prod_{l\in S_{k,j}}d\theta_l}{\prod_{kl\in S_{k,j}} \int_{\theta_l\in[0,2\pi]}d\theta_l}$$
where
$$\tilde{a}_k(r):= (r_{k_1}^2+r_{k_2}^2+r_{k_3}^2-r_{k_4}^2-r_{k_5}^2-r_{k_6}^2).$$
 The only terms different from 0 are the terms where $$\theta_{j_1}+\theta_{j_2}+\theta_{j_3}+\theta_{k_4}+\theta_{k_5}+\theta_{k_6}=\theta_{j_4}+\theta_{j_5}+\theta_{j_6}+\theta_{k_1}+\theta_{k_2}+\theta_{k_3}$$
or equivalently
 $$\left\{j_1,j_2,j_3,k_4,k_5,k_6\right\}= \left\{j_4,j_5,j_6,k_1,k_2,k_3\right\}.
$$

This implies that the integrals that survive have this form:
$$\frac{\int_{r_k\in\mathbb{R}_+} r_{j_1}^2r_{j_2}^2r_{j_3}^2{r}_{k_4}^2{r}_{k_5}^2{r}_{k_6}^2\left(1-\rho\left(\frac{\tilde{a}_j(r)}{\delta}\right) \right)\left(1-\rho\left(\frac{\tilde{a}_k(r)}{\delta}\right) \right)e^{-\beta\sum_{l\in S_{k,j}}\left(1+ l^2\right)r_l^2}\prod_{l\in S_{k,j}} r_ldr_l}{\prod_{l\in S_{k,j}}\int_{\mathbb{R}_+}e^{-\beta\left(1+ l^2\right)r_l^2} r_ldr_l}=$$

$$\frac{\int_{z_k\in\mathbb{R}_+} z_{j_1}z_{j_2}z_{j_3}{z}_{k_4}{z}_{k_5}{z}_{k_6}\left(1-\rho\left(\frac{\tilde{b}_j(z)}{\beta\delta}\right) \right)\left(1-\rho\left(\frac{\tilde{b}_k(z)}{\beta\delta}\right) \right)e^{-\sum_{l\in S_{k,j}}{z_l}}\prod_{l\in S_{k,j}}dz_l}{\beta^6{\left(1+j_1\right)}^2{\left(1+j_2\right)}^2{\left(1+j_3\right)}^2{\left(1+k_4\right)}^2{\left(1+k_5\right)}^2{\left(1+k_6\right)}^2\prod_{l\in S_{k,j}}\int_{\mathbb{R}_+}e^{-\sum_l{z_l}}dz_l}$$
where
$$\tilde{b}_k(z):= \left(\frac{z_{k_1}}{1+k_1^2}+\frac{z_{k_2}}{1+k_2^2}+\frac{z_{k_3}}{1+k_3^2}-\frac{z_{k_4}}{1+k_4^2}-\frac{z_{k_5}}{1+k_5^2}-\frac{z_{k_6}}{1+k_6^2}\right).$$

We define the non smooth cutoff function $\chi(x)=
\left\{
\begin{array}{ll}
0 \mbox{ if } |x|\geq \delta\beta \\
1 \mbox{ if } |x|
\leq \delta\beta
\end{array}
\right.
$

So we can increase the integral with the following integral:
$$\frac{1}{\beta^6{\left(1+j_1\right)}^2{\left(1+j_2\right)}^2{\left(1+j_3\right)}^2{\left(1+k_4\right)}^2{\left(1+k_5\right)}^2{\left(1+k_6\right)}^2}\times$$
\begin{equation}\label{intres}
\int \prod_{i=1}^3z_{j_i}\prod_{l=4}^6{z}_{k_l}\chi\left(\tilde{b}_j(z)\right)\chi\left(\tilde{b}_k(z)\right)e^{-\sum_{l\in S_{k,j}}{z_l}}\prod_{l\in S_{k,j}} dz_l.
\end{equation}
We would to know more information on the arguments of the  cutoff function  that depend on the form of $Z_{6,k,\tt k}$ and $Z_{6,j,\tt k}$.  
\\Since  in $\cR_6^{R}$  there are only terms in which  $\left\{k_1,k_2,k_3\right\}\neq \left\{k_4,k_5,k_6\right\}$, this implies also that there are only terms  in which $k_i\neq k_l$ for $i=1,2,3$ $l=4,5,6$, since if there  exists at least an index $ i\in \{1,2,3\},$ and index $ l\in \{4,5,6\}$ s.t. $k_i=k_l$ this implies that $\left\{k_1,k_2,k_3\right\}= \left\{k_4,k_5,k_6\right\}$ and it is absurd.

In fact, without losing generality we can  suppose that  $k_1=k_4$, this means that $k_2+k_3=k_5+k_6$ and $k_2^2+k_3^2=k_5^2+k_6^2$, so $k_2=k_5$ and $k_3=k_6$ or $k_2=k_6$ and $k_3=k_5$, so $\left\{k_1,k_2,k_3\right\}= \left\{k_4,k_5,k_6\right\}$.

So one has $j_i\neq j_l$ and $k_i\neq k_l$ $j=1,2,3$, $l=4,5,6$. Moreover we know that $\left\{j_1,j_2,j_3,k_4,k_5,k_6\right\}= \left\{j_4,j_5,j_6,k_1,k_2,k_3\right\}$ this means $\left\{j_1,j_2,j_3\right\}= \left\{k_1,k_2,k_3\right\}$ and $\left\{k_4,k_5,k_6\right\}= \left\{j_4,j_5,j_6\right\}$ and $\left\{j_1,j_2,j_3,j_4,j_5,j_6\right\}=\left\{k_1,k_2,k_3,k_4,k_5,k_6\right\}=\left\{j_1,j_2,j_3,k_4,k_5,k_6\right\}$

So,  up to any permutation of the indices, we have 9 cases:
\begin{itemize}
\item if $j_i\neq j_l$, $k_i\neq k_l$, $\tilde{b}_k(z)=\tilde{b}_j(z)= \left(\frac{z_{j_1}}{1+j_1^2}+\frac{z_{j_2}}{1+j_2^2}+\frac{z_{j_3}}{1+j_3^2}-\frac{z_{k_4}}{1+k_4^2}-\frac{z_{k_5}}{1+k_5^2}-\frac{z_{k_6}}{1+k_6^2}\right),$
\item if $j_i\neq j_l$, $k_4=k_5$, $\tilde{b}_k(z)=\tilde{b}_j(z)= \left(\frac{z_{j_1}}{1+j_1^2}+\frac{z_{j_2}}{1+j_2^2}+\frac{z_{j_3}}{1+j_3^2}-2\frac{z_{k_4}}{1+k_4^2}-\frac{z_{k_6}}{1+k_6^2}\right),$
\item if $j_i\neq j_l$, $k_4=k_5=k_6$, $\tilde{b}_k(z)=\tilde{b}_j(z)= \left(\frac{z_{j_1}}{1+j_1^2}+\frac{z_{j_2}}{1+j_2^2}+\frac{z_{j_3}}{1+j_3^2}-3\frac{z_{k_4}}{1+k_4^2}\right),$
\item if $j_1=j_2$, $k_i\neq k_l$, $\tilde{b}_k(z)=\tilde{b}_j(z)= \left(\frac{2z_{j_1}}{1+j_1^2}+\frac{z_{j_3}}{1+j_3^2}-\frac{z_{k_4}}{1+k_4^2}-\frac{z_{k_5}}{1+k_5^2}-\frac{z_{k_6}}{1+k_6^2}\right),$
\item if $j_1=j_2$,  $k_4=k_5$, $\tilde{b}_k(z)=\tilde{b}_j(z)= \left(\frac{2z_{j_1}}{1+j_1^2}+\frac{z_{j_3}}{1+j_3^2}-2\frac{z_{k_4}}{1+k_4^2}-\frac{z_{k_6}}{1+k_6^2}\right),$
\item if $j_1=j_2$, $k_4=k_5=k_6$, $\tilde{b}_k(z)=\tilde{b}_j(z)= \left(\frac{2z_{j_1}}{1+j_1^2}+\frac{z_{j_3}}{1+j_3^2}-3\frac{z_{k_4}}{1+k_4^2}\right),$
\item if $j_1=j_2=j_3$, $k_i\neq k_l$, $\tilde{b}_k(z)=\tilde{b}_j(z)= \left(3\frac{z_{j_1}}{1+j_1^2}-\frac{z_{k_4}}{1+k_4^2}-\frac{z_{k_5}}{1+k_5^2}-\frac{z_{k_6}}{1+k_6^2}\right),$
\item if $j_1=j_2=j_3$, $k_4=k_5$, $\tilde{b}_k(z)=\tilde{b}_j(z)= \left(3\frac{z_{j_1}}{1+j_1^2}-2\frac{z_{k_4}}{1+k_4^2}-\frac{z_{k_6}}{1+k_6^2}\right),$
\item if $j_1=j_2=j_3$, $k_4=k_5=k_6$, $\tilde{b}_k(z)=\tilde{b}_j(z)= \left(3\frac{z_{j_1}}{1+j_1^2}-3\frac{z_{k_4}}{1+k_4^2}\right).$
\end{itemize}
We can resume  all this cases writing $$\tilde{b}_k(z)=\tilde{b}_j(z)=\tilde{b}_{kj}(z)=$$
$$= \left(a_1\frac{z_{j_1}}{j_1^2}+a_2\frac{z_{j_2}}{1+j_2^2}+a_3\frac{z_{j_3}}{1+j_3^2}-a_4\frac{z_{k_4}}{1+k_4^2}-a_5\frac{z_{k_5}}{1+k_5^2}-a_6\frac{z_{k_6}}{1+k_6^2}\right)$$
 where  $a_i\in\left\{0,1,2,3\right\}$, $\sum_{i=1}^6a_i= 6$, and $\{a_i\}_{i=1}^6$ s.t. if there exists $i\in\{1,2,3\}$ s.t. $a_i\neq 1$, for any $l\in\{1,2,3\}$, $l\neq i$ s.t. $a_l=0$, $j_i=j_l$ and if there exists $i'\in\{4,5,6\}$ s.t $a_{i'}\neq 1$, for any $l'\in\{4,5,6\}$, $l'\neq i'$ s.t. $a_{l'}=0$, $k_{i'}=k_{l'}$.
In this way we can write 
$\eqref{intres}$ as 
\begin{equation}\label{intres2}
\frac{1}{\beta^6\prod_{i=1}^3\left(1+j^2_i\right)\left(1+k^2_{3+i}\right)}\int \prod_{i=1}^3z_{j_i}{z}_{k_{3+i}}
\chi\left(\tilde{b}_{kj}(z)\right)e^{-\sum_{l\in S_{k,j}} z_l}\prod_{l\in S_{k,j}} dz_l
\end{equation}
where $z_i\in \mathbb{R}_+$.

To obtain the norm of the  resonant part, after studying the form of any terms of the series, we have to estimate the norm of every single term.

Let $N$ be an integer,  then Lemma $\ref{piccoli}$ shows that if there exists at least an index $i=1,2,3$, $a_i\neq0$ s.t. $|j_i|<N$  or an index $l=4,5,6$, $a_l\neq0$ s.t. $|k_l|<N,$ then there exists $C_1>0$ s.t.  \eqref{intres2} is bounded by

$$C_1\frac{\delta\beta N^2}{\prod_{i=1}^3\left(1+j^2_i\right)\left(1+k^2_{3+i}\right)}.$$

If every $j_i$ and $k_l$ really present in the argument of the cutoff is bigger than $N$, we adopt an  other strategy, because the distance between the two hyper-planes becomes bigger and non comparable with $\delta\beta$, so the presence of the cutoff isn't so essential, because the integral isn't so different from the integral over  all the space. However, if  all the indices in the argument of the cutoff  are bigger than $N$, the denominators  $\beta^6\prod_{i=1}^3\left(1+j^2_i\right)\left(1+k^2_{3+i}\right)$ is small and this helps the convergence.
Obviously, since there exists at least an index $j_i$ or $k_i$ equal to $\tt k$, this situation is possible only if $|{\tt k}|\geq N$.

We denote by $T_{\tt k}$ the set of $(k,j)\in \mathbb{Z}^{12}$ s.t. $\{j_1,j_2,j_3,k_4,k_5,k_6\}=\{k_1,k_2,k_3,j_4,j_5,j_6\}$, $\sum_{i=1}^nk_i=\sum_{i=n+1}^{2n}k_i$, $\sum_{i=1}^nj_i=\sum_{i=n+1}^{2n}j_i$, and s.t. there exists at least an index $i\in\{1,2,3,4,5,6\}$ s.t. $k_i=\tt k$ and at least an index $l\in\{1,2,3,4,5,6\}$ s.t. $j_l=\tt k$.

So, if  ${\tt k}<N$, we have $$\left\|\cR_6^{R}\right\|^2_g\leq 9C_1\frac{\delta\beta N^2}{\beta^6}\sum_{\substack{j,k\in T_{\tt k} }}\frac{|\tilde{Z}_{6,j}||\tilde{Z}_{6,k}|}{\prod_{i=1}^3\left(1+j^2_i\right)\left(1+k^2_{3+i}\right)}.$$
 Instead, if  ${\tt k}\geq N$, we have  that $\left\|\cR_6^{R}\right\|^2_g$ is bounded by
$$  9C_1\frac{\delta\beta N^2}{\beta^6}\sum_{\substack{j,k\in T_{\tt k} }}\frac{|\tilde{Z}_{6,j}||\tilde{Z}_{6,k}|}{\prod_{i=1}^3\left(1+j^2_i\right)\left(1+k^2_{3+i}\right)}$$
$$+ \frac{9}{\beta^6}\sum_{\substack{j,k\in T_{\tt k} \mbox{ s.t }\\  \forall i\; |j_i|,|k_i|\geq N}}\frac{|\tilde{Z}_{6,j}||\tilde{Z}_{6,k}|}{\prod_{i=1}^3\left(1+j^2_i\right)\left(1+k^2_{3+i}\right)}.$$
We know also that for every $j$ in the sum  there is an index $i$ s.t. $j_i=\tt k$ but, due to the \textit{null momentum condition}, there must be at  least an other index $l$ s.t. $|j_l|\geq \frac{|\tt k|}{5}$ and the same holds also for any $k$. Moreover, from Lemma $\ref{parentesi_poisson_lemma}$,  $|\tilde{Z}_{6,j}|$ are uniformly limited by a constant. So, in both  the cases, as in Theorem $\ref{gau}$,  we have
$$\sum_{\substack{j,k\in T_{\tt k} }}\frac{|\tilde{Z}_{6,j}||\tilde{Z}_{6,k}|}{\prod_{i=1}^3\left(1+j^2_i\right)\left(1+k^2_{3+i}\right)}\leq\frac{C}{\left(1+k^2\right)^2}\sum_{l_1,l_2,l_3,l_4}\frac{1}{\prod_{i=1}^4{\left(1+l_i^2\right)}}$$ 
and, choosing $0<\epsilon\ll1$, 
$$\sum_{\substack{j,k\in T_{\tt k} \mbox{ s.t }\\  \forall i\; |j_i|,|k_i|\geq N}}\frac{|\tilde{Z}_{6,j}||\tilde{Z}_{6,k}|}{\prod_{i=1}^3\left(1+j^2_i\right)\left(1+k^2_{3+i}\right)}\leq\frac{C}{\left(1+k^2\right)^2}\sum_{\substack{l_1,l_2,l_3,l_4\\\forall i,\; |l_i|>N}}\frac{1}{\prod_{i=1}^4{\left(1+l_i^2\right)}}$$$$\leq\frac{C}{\left(1+k^2\right)^2N^{4-4\epsilon}}\sum_{\substack{l_1,l_2,l_3,l_4\\\forall i,\; |l_i|>N}}\frac{1}{\prod_{i=1}^4{\left(1+l_i^2\right)^{\frac{1+\epsilon}{2}}}}.$$
One has $\sum_{\substack{l_1,l_2,l_3,l_4\\\forall i,\; |l_i|>N}}\frac{1}{\prod_{i=1}^4{\left(1+l_i^2\right)^{\frac{1+\epsilon}{2}}}}\sim \frac{1}{N^{4\epsilon}}$, so,
 we can take $$\delta\beta N^2=\frac{1}{N^{4}},$$ one has $N=\frac{1}{(\delta\beta )^{\frac{1}{6}}}$ and finally 
$$\delta\beta N^2=\frac{1}{N^{4}}=(\delta\beta)^{\frac{2}{3}}.$$
This implies that $$\left\|\cR_6^{R}\right\|^2_g\leq \tilde{C}\frac{(\delta\beta)^{\frac{2}{3}}}{\beta^6\left(1+k^2\right)^2}.$$
\qed


\bibliography{azioni_1}

\end{document}